\documentclass[twocolumn]{aastex62}

\usepackage{graphicx}	% Including figure files
\usepackage{amsmath}	% Advanced maths commands
\usepackage{amssymb}	% Extra maths symbols
\usepackage{hyperref}
\usepackage{natbib}

\graphicspath{{./}{images/}}

\received{April 11, 2019}
\revised{May 23, 2019}
\accepted{\today}
\submitjournal{ApJ}

\shorttitle{CWB in M33}
\shortauthors{Garofali et al.}

\begin{document}
\title{The First Candidate Colliding-Wind Binary in M33}

\correspondingauthor{Kristen Garofali}
\email{garofali@uark.edu}

\author[0000-0002-9202-8689]{Kristen Garofali}
\affil{Department of Physics, University of Arkansas, 825 West Dickson St, Fayetteville, AR 72701}
 
 \author{Emily M. Levesque}
 \affil{Department of Astronomy, University of Washington, Box 351580, Seattle, WA 98195}
 
    \author{Philip Massey}
 \affil{Lowell Observatory, 1400 W Mars Hill Road, Flagstaff, AZ 86001} 
  \affil{Department of Physics and Astronomy, Northern Arizona University, Flagstaff, AZ, 86011-6010}

 \author{Benjamin F. Williams}
 \affil{Department of Astronomy, University of Washington, Box 351580, Seattle, WA 98195}

\begin{abstract}
%250 words
We present the detection of the first candidate colliding-wind binary (CWB) in M33, located in the giant H II region NGC 604. The source was first identified in archival {\it Chandra} imaging as a relatively soft X-ray point source, with the likely primary star determined from precise astrometric alignment between archival {\it Hubble Space Telescope} and {\it Chandra} imaging. The candidate primary star in the CWB is classified for the first time in this work as a carbon-rich Wolf-Rayet star with a likely O star companion based on spectroscopy obtained from Gemini-North. We model the X-ray spectrum using {\it Chandra} and {\it XMM-Newton} observations, and find the CWB is well-fit as a $\sim$ 1 keV thermal plasma with a median unabsorbed luminosity in the 0.5--2.0 keV band of $L_{\rm X}$ $\sim$ 3 $\times$ 10$^{35}$ erg s$^{-1}$, making this source among the brightest of CWBs observed to date. We present a long term light curve for the candidate CWB from archival {\it Chandra} and {\it XMM-Newton} observations, and discuss the constraints placed on the binary by this light curve, as well as the X-ray luminosity at maximum. Finally, we compare this candidate CWB in M33 to other well-studied, bright CWBs in the Galaxy and Magellanic Clouds, such as $\eta$ Car. 
\end{abstract}


%%%%%%%%%%%%%%%%%%%%%%%%%%%%%%%%%%%%%%%%%%%%%%%%%%

%%%%%%%%%%%%%%%%% BODY OF PAPER %%%%%%%%%%%%%%%%%%

\section{Introduction}

Massive stars, stars $\gtrsim$ 8 $M_{\odot}$, vigorously shape their surroundings via mechanical, kinetic, and chemical feedback driven by their strong radiative winds and explosive deaths. The mass loss history of a massive star not only shapes its stellar surroundings, but also significantly alters the evolution of the star itself. A high fraction of massive stars have also been found to exist in close binaries which, if they do not merge, will either accrete or strip mass from their companion at some point during their evolution \citep{Sana2012,Neugent2014}. Thus mass loss via stellar winds or interactions in a binary are of paramount importance for understanding massive star evolution and its influence via energetic and chemical feedback.

Theoretical models of wind mass loss rates \citep[e.g.,][]{Vink2001} have long been used as input in stellar evolution codes to predict evolution with mass loss history; however, mass loss rate predictions from theoretical models can be discrepant from empirically derived values by a factor of a few or more \citep{Fullerton2006,Oskinova2007,Sundqvist2011,Bouret2012}. The systematic biases in observed mass loss rates depend on choice of line diagnostic, as $\rho^{2}$ diagnostics such as H$\alpha$ are especially sensitive to wind ``clumpiness" \citep{Kud2000,Smith2014}. This highlights the need for a multiwavelength approach to empirically constraining massive star wind properties \citep[e.g.,][]{Cohen2011,Cohen2014,Puebla2016}, and therefore the effects of mass loss on massive star evolution and the surrounding environment. 

To this end, X-ray emission from massive stars provides a means to probe the structure and strength of stellar winds beyond standard UV, IR, and optical diagnostics, which are more sensitive to the systematics associated with non-uniform, or clumpy winds \citep{Pittard2007}. In particular, X-ray line diagnostics and X-ray variability from both single stars and binary systems can be used to probe wind strength and structure \citep[e.g.,][]{Pittard1997,Pittard2002,Oskinova2004,Naze2013}. 

Single OB stars produce soft ($\lesssim$ keV) X-ray emission with characteristic $L_{\rm X}$ $\sim$ 10$^{31}$--10$^{33}$ erg s$^{-1}$ from shocks embedded in their stellar winds \citep{Lucy1970,Lucy1980,Lucy1982,Owocki1988,Feld1997}, and in addition clusters of hot, young stars can drive wind-blown bubbles that themselves are sources of soft, diffuse X-ray emission \citep[e.g.,][]{Tullmann2008}. Slightly harder (T $>$ 10$^{7}$ K) X-ray emission with characteristic $L_{\rm X}$ $\sim$ 10$^{32}$--10$^{35}$ erg s$^{-1}$ can be produced by shocks originating in wind-wind collision regions of massive binaries (O + Wolf-Rayet [WR], or WR + WR), sources known as colliding-wind binaries (CWBs) \citep{Pri1976,Cher1976,Pollock1987,Luo1990,Stevens1992,Usov1992}. 

While both single X-ray emitting massive stars and CWBs can be used to probe the properties of massive star winds \citep[e.g.,][]{Stevens1996,Oskinova2003,Pittard2006,Lomax2015} CWBs are on average more X-ray luminous than single stars \citep{Pollock1987,Clark2008,Guerrero2008}, and as such act as some of the brightest beacons in X-rays in the Galaxy and Magellanic Clouds (MCs) for studying in detail the strength and structure of winds from massive stars. However, the brightest CWBs such as $\eta$ Carinae, which reach $L_{\rm X}$ $\sim$ 10$^{35}$ erg s$^{-1}$, are somewhat rare \citep{Gudel2009,Naze2011}, thus limiting the most detailed studies of such systems to perhaps a dozen sources in Galaxy and MCs \citep[e.g.,][]{Rauw2016}, with none yet observed beyond the MCs. 

Here we report the discovery of the first candidate CWB in M33 located in the giant H II region NGC 604. The system, hereafter referred to N604-WRX, hosts a carbon-rich WR (WC) primary star, and is comparable in luminosity to the most X-ray bright CWBs in the Galaxy and MCs \citep[$L_{x}$ $\sim$ 10$^{35}$ erg s$^{-1}$;][]{Pollock2018}. 

This paper is organized as follows: in Section~\ref{sec:reduce} we describe the X-ray and optical observations used to identify and characterize N604-WRX. In Section~\ref{sec:results} we present our model for the X-ray spectrum of N604-WRX, and the X-ray luminosity and light curve derived from this model. We also present the spectral classification of the likely primary star in the CWB using ground-based spectroscopy. Next, in Section~\ref{sec:discuss} we discuss the evidence from our observations in favor of N604-WRX as a CWB, and compare the source properties to other bright CWBs in the Galaxy and MCs. Finally, we provide conclusions and a discussion of future work to characterize N604-WRX in Section~\ref{sec:conclude}. 

\begin{deluxetable*}{ccccccc}
\tablewidth{\textwidth}
 \tablecaption{Archival {\it Chandra} observations used in this work. Column (1): observation start date, column (2): observation ID, columns (3)--(4): instrument and observing mode, column (5): off-axis angle in arcminutes of N604-WRX in each observation, and column (6): effective exposure times in ks after GTI correction.\label{tab:chobs}}
\tablehead{
 \colhead{Obs. Start Date} & \colhead{Obs. ID} & \colhead{Instrument} & \colhead{Obs. Mode} & \colhead{Off-Axis Angle} & \colhead{Eff. Exposure} & \colhead{PI} \\ 
 \colhead{} & \colhead{} &\colhead{} & \colhead{} & \colhead{(arcmin)} & \colhead{(ks)} & \colhead{} } 

\startdata 
2001-07-06 & 2023 & ACIS-I & FAINT & 1.0 & 69.4  & F. Damiani \\
2005-09-21 & 6378 & ACIS-I & VFAINT & 4.28 & 83.1 & M. Sasaki \\ 
2006-06-09 & 6388 & ACIS-I & VFAINT & 8.06 & 80.1  & M. Sasaki \\
2006-09-04 & 6379 & ACIS-I & VFAINT & 4.23 & 48.8  & M. Sasaki \\ 
2006-09-07 & 7402 & ACIS-I & VFAINT & 4.22 & 39.4 & M. Sasaki \\
\enddata 
\end{deluxetable*}

\section{Observations \& Data Reduction}\label{sec:reduce}

In this section, we present an overview of the data products, reduction, and analysis methods used in characterizing the candidate CWB N604-WRX, namely archival {\it Chandra}, {\it XMM-Newton}, and {\it Hubble Space Telescope (HST)} imaging, as well as spectra of the likely primary star newly obtained with Gemini-North.  

\subsection{{\itshape Chandra} Imaging \& Spectra}\label{sec:chobs}

We use five archival {\it Chandra} observations to characterize X-ray emission from N604-WRX. Four observations (obs. IDs 6378, 6379, 6388, and 7204) come from the deep {\it Chandra} ACIS survey of M33 \citep[ChASeM33;][]{Tullmann2011} and were taken with ACIS-I in ``Very Faint" (VFAINT) mode. One archival observation (obs. ID 2023) was taken with NGC 604 as the target, with the observations performed with ACIS-I in ``FAINT" mode. The observations are summarized in Table~\ref{tab:chobs}, including off-axis angle of N604-WRX in each exposure, and effective exposure time for each observation after filtering on good time intervals (GTIs).

We reduced all archival observations using standard reduction techniques with {\tt CIAO} version 4.10  and {\tt CALDB} version 4.8.1\footnote{\url{http://cxc.harvard.edu/ciao/download/}}. We reprocessed the {\tt level=1} event files using the script {\tt chandra$\_$repro}. Subsequently, we produced background light curves for each observation, where time intervals with high count rates were removed by filtering the {\tt level=2} event files on the GTIs as determined from the task {\tt lc$\_$clean} with 5$\sigma$ clipping. The effective exposures for each observation after filtering for background flaring are listed in Table~\ref{tab:chobs}, resulting in a total effective exposure for N604-WRX of  $\sim$ 320 ks. 

Next, we ran {\tt wavdetect} independently on each exposure to determine source positions for matching between exposures. We fixed the position for N604-WRX to the position returned from {\tt wavdetect} from the most on-axis observation (obs. ID 2023), finding good agreement between this position (RA: 1:34:32.606, Dec: +30:47:04.091) and the corresponding point source position (013432.60+304704.1) listed in the ChASeM33 catalog \citep{Tullmann2011}. We then ran {\tt wcs$\_$match} between all other exposures and obs. ID 2023 using sources returned from {\tt wavdetect}, and subsequently updated the aspect solutions and event files for each observation using {\tt wcs$\_$update} to bring them into alignment with obs. ID 2023. 

\begin{figure}
\centering
\includegraphics[width=0.5\textwidth,trim=0 0 0 0, clip]{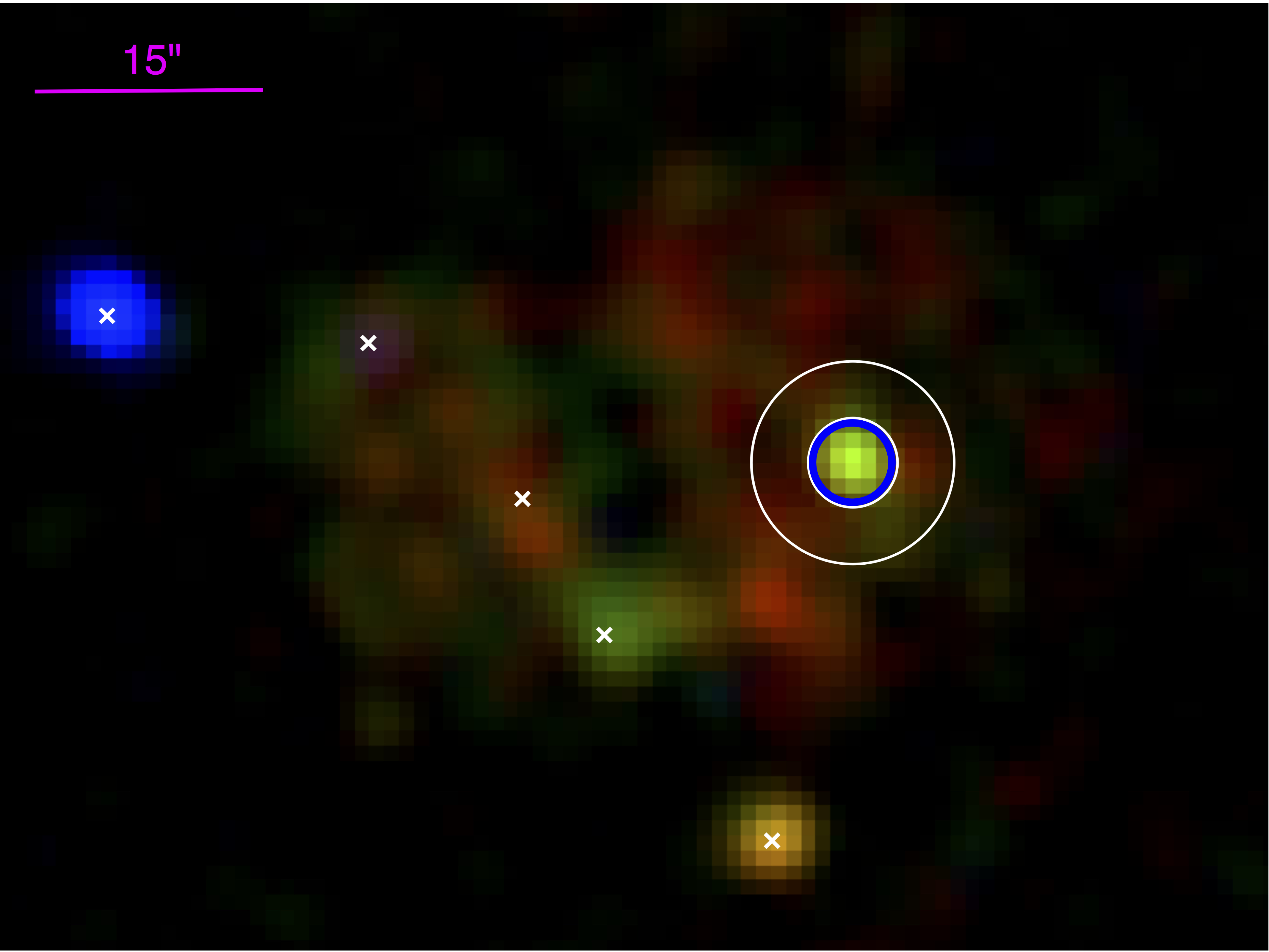}
\caption{Three-color ({\it red}: 0.2-1.0 keV, {\it green}: 1.0-2.0 keV, {\it blue}: 2.0-8.0 keV), exposure-corrected mosaic of NGC 604, binned into 1" $\times$ 1" pixels, and smoothed using a Gaussian filter with $\sigma$=3. North is up and east is to the left. The source extraction region for N604-WRX is overlaid in blue just inside the background extraction annulus (white). Other X-ray sources in NGC 604 from \citet{Tullmann2011} are marked as white x's on the image: notably, a hard source to the NE, and a bright supernova remnant south of NGC 604.}\label{fig:chandra_rgb}
\end{figure}

With the position of N604-WRX determined for all exposures, we proceeded to extract spectral products. For each exposure we determined the region encircling 95\% of the point spread function (PSF) of the source using the task {\tt psfsize$\_$srcs}, and used these regions for source extraction. We also extracted background regions in order to model the surrounding soft, diffuse emission from NGC 604. The background regions were chosen as annuli surrounding the source extraction regions, with the inner radius set to where the radial surface brightness profile of N604-WRX approached the level of emission from NGC 604 in the surrounding region, and the outer radius set so that the background annulus was roughly five times the size of the source extraction region, ensuring enough counts to adequately characterize the background. Using these source and background regions, we extracted spectral products using the task {\tt specextract} for each exposure: source and background spectra, response matrix files (RMFs), and auxiliary response files (ARFs). Details of the spectral models and fitting process and discussed in Section~\ref{sec:xspec}. 

Finally, we created a merged, exposure-corrected image using all obs. IDs listed in Table~\ref{tab:chobs}. The individual exposures were first reprojected onto the frame of obs. ID 2023 and then merged using the {\tt CIAO} task {\tt reproject$\_$obs}. The merged image was then exposure corrected and binned into $\sim$ 1" $\times$ 1" pixels (2 $\times$ 2 binning) using the task {\tt flux$\_$obs}. As the data were processed using the Energy-Dependent Subpixel Event Repositioning algorithm \citep[EDSER;][]{Li2004} during the {\tt chandra$\_$repro} task, we also created an image attempting to recover sub-pixel resolution for N604-WRX using an image of the source PSF produced with {\tt MARX}, and the task {\tt dmcopy} imposing a binning factor less than one; however, this process appeared to generate artifacts not present in the image at native resolution. Thus, we opted to use the merged image with 1" $\times$ 1" pixels to create exposure-corrected images in the soft (0.2-1.0 kev), medium (1.0-2.0 keV), and hard (2.0-8.0 keV) bands. In Figure~\ref{fig:chandra_rgb} we show the three-color, exposure-corrected mosaic of NGC 604 created using these images with source (blue) and background (white) spectral extraction regions for N604-WRX overlaid. N604-WRX is clearly visible as a point source in the western region of NGC 604 with counts in both the soft and medium bands. 

\begin{deluxetable*}{ccccccccc}
\tablewidth{1.0\textwidth}
 \tablecaption{Archival {\it XMM-Newton} observations used in this work. Column (1): observation start date, column (2): observation ID, columns (3)--(5): off-axis angle in arcminutes of N604-WRX in each exposure, and columns (6)--(8): effective exposure time in ks after GTI correction. Exposures without listed off-axis angles or effective exposures times are observations in which N604-WRX was off-field or on a chip gap. \label{tab:xmmobs}}
\tablehead{
 \colhead{Obs. Start Date} & \colhead{Obs. ID} & \multicolumn{3}{|c|}{Off-Axis Angle (arcmin)} & \multicolumn{3}{|c|}{Eff. Exposure (ks)} & \colhead{PI} \\ 
 \colhead{} & \colhead{} & \colhead{MOS1} & \colhead{MOS2} & \colhead{PN} & \colhead{MOS1} & \colhead{MOS2} & \colhead{PN} & \colhead{}}

 \startdata 
2010-07-09 & 0650510101 & 4.8 & 5.7 & 4.1 & 87.0 & 74.2 & 81.2 & B.F. Williams \\
2010-07-21 & 0650510301 & --  & -- & 15.7 & -- & --  & 79.4 & B.F. Williams \\ 
2012-01-10 & 0672190301 & -- & 8.5 & 7.5 & -- & 97.6 & 93.4 & B.F. Williams \\ 
2017-08-02 & 0800350301 & -- & 5.4 & -- & -- & 23.3 & -- & B. Lehmer \\
\enddata 
\end{deluxetable*}

\newpage

\subsection{{\itshape XMM-Newton} Imaging \& Spectra}

To further characterize the X-ray luminosity and spectrum of N604-WRX, we downloaded all available {\it XMM-Newton} observations in the vicinity of the source from the {\it XMM-Newton} Science Archive (XSA)\footnote{\url{http://nxsa.esac.esa.int/nxsa-web/}}, comprising four separate observations and a total of seven exposures where N604-WRX was not off-field or on a chip gap. The observations are summarized in Table~\ref{tab:xmmobs}.

Observational data files (ODFs) for each observation were reprocessed using the {\it XMM-Newton} Science Analysis System ({\tt SAS} version 17.0)\footnote{\url{https://www.cosmos.esa.int/web/xmm-newton/sas}}. To produce processed event lists from the ODF for the EPIC-MOS \citep{Turner2001} and EPIC-pn detectors \citep{Struder2001} we employed the {\tt SAS} tasks {\tt emchain} and {\tt epchain}, respectively, applying standard filtering as follows: (PATTERN $\leq$ 12 \&\& flag== \#XMMEA$\_$EM) for the EPIC-MOS detectors to include single, double, and quadruple events and standard event flagging, and (PATTERN $\leq$ 4 \&\& FLAG == 0) for the EPIC-pn detector to include single and double events and conservative flagging. 

With these event lists for each exposure, we next created field-wide high-energy light curves to determine GTIs to filter the event lists for background flaring. For obs. IDs 0650510101, 0650510301, and 0672190301 we created 7-15 keV light curves for all detectors as in \citet{Williams2015}. We inspected each light curve by eye to determine the periods of high count rate to exclude:  $>$ 0.35 counts s$^{-1}$ for 0650510101 and 0672190301 MOS detectors, $>$ 3 counts s$^{-1}$ for 0650510101 and 0672190301 pn,  $>$ 0.5 counts s$^{-1}$ for 0650510301 MOS detectors, and $>$ 17.5 counts s$^{-1}$ for 0650510301 pn. For obs. ID 0800350301 we created $>$ 10 keV light curves for the MOS detectors and 10-12 keV light curves for the pn, filtering out periods with count rates $>$ 0.35 counts s$^{-1}$ for the MOS, and 0.4 counts s$^{-1}$ for the pn as in \citet{West2018}. The effective exposures for each exposure after GTI correction are listed in Table ~\ref{tab:xmmobs}, resulting in a total effective exposure for N604-WRX of $\sim$ 536 ks. 

After GTI-correction, we performed source detection using {\tt edetect$\_$chain} separately for each exposure to determine source positions for cross-correlation with the positions from {\it Chandra} obs. ID 2023. Translational shifts between each {\it XMM-Newton} exposure and {\it Chandra} were determined using at least three sources per exposure with the {\tt CIAO} task {\tt wcs$\_$match}, and subsequently used to bring each EPIC exposure into alignment with {\it Chandra}. Given the translational shifts returned for each EPIC exposure, we estimate an additional positional uncertainty for N604-WRX of 2\farcs1 in RA and 0\farcs75 in Dec. for all EPIC exposures, which we take into account during source spectral extraction. Using the translational shifts from cross-correlation with {\it Chandra} we determine the position of N604-WRX in each EPIC exposure for use in source spectral extraction, as described below. 

As for the {\it Chandra} observations, spectral extraction region sizes for N604-WRX in each EPIC exposure were chosen to encompass 95\% of the PSF at energies of 1 keV, as determined using the {\tt SAS} tasks {\tt psfgen} and {\tt calview} \citep{Read2011} and the off-axis angle of N604-WRX in each exposure. In addition, we include the positional uncertainty due to alignment in the extraction region sizes. To characterize the sky and instrumental background we selected elliptical regions twice the size of the source extraction regions, which for PN were at similar RAWY positions and off-axis angles as N604-WRX, and for MOS were on the same CCD and at a similar off-axis angle as N604-WRX. For all exposures we ensured that the source and background regions were free of other point sources, both through by-eye verification and checks against other point source catalogs \citep{Tullmann2011,Williams2015}. 
 
We next extracted source and background spectra for N604-WRX using {\tt evselect} with a spectral bin size of five for the PN, and 15 for MOS exposures. RMFs and ARFs were produced using the tasks {\tt rmfgen} and {\tt arfgen}. In Table~\ref{tab:xmmobs} we list all exposures for which spectral products were extracted. Those exposures without off-axis angles or effective exposures listed are exposures in which N604-WRX was either off-field or on a chip gap.

After inspection of the reduced {\it XMM-Newton} imaging, we find that the point source N604-WRX is not distinguishable from the surrounding soft emission in NGC 604. This is likely due to the higher soft-sensitivity and lower angular resolution of {\it XMM-Newton} as compared to {\it Chandra}. Given our inability to resolve N604-WRX from the background in {\it XMM-Newton}, we use the archival {\it XMM-Newton} observations only to place upper limits on the flux of N604-WRX as described in Section~\ref{sec:xspec}. 

\subsection{{\itshape HST} Photometry}

In addition to extracting all available X-ray observations of N604-WRX, we searched for the optical counterpart to the X-ray emission to identify the most likely primary star in the CWB. Given the crowded nature of the field, with $\sim$ 200 OB stars in the western region of NGC 604 alone \citep{Hunter1996}, the high angular resolution of {\it Chandra} and {\it HST} were necessary to identify possible optical counterparts to N604-WRX within the X-ray positional uncertainty. Below, we describe the {\it HST} observations used to make the initial identification of the candidate optical counterpart, or primary star, for N604-WRX. 

The giant H II region NGC 604 was first imaged with {\it HST} using the Wide Field Planetary Camera 2 (WFPC2) in two filters (F555W and F814W), each with exposure times of 400 s (proposal ID 5237, field M33-FIELDN604). The data for this field were reduced as described in \citet{Garofali2018}, using the pipelines developed as part of the ACS Nearby Treasury Program \citep[ANGST;][]{Dalcanton2009}, and the Panchromatic Hubble Andromeda Treasury program \citep[PHAT;][]{Dalcanton2012}. Photometry was measured using {\tt DOLPHOT} and {\tt HSTPHOT} \citep{Dolphin2000}, and completeness was estimated using artificial star tests as described in \citet{Williams2014} and \citet{Garofali2018}. The 50\% completeness limits in this field are $m_{\rm F555W}^{50\%}$ = 24.33, and $m_{\rm F814W}^{50\%}$ = 24.21, roughly the magnitude of a B star at the distance of M33. 

The WFPC2 image was aligned to a common frame as the {\it Chandra} observations, again as described in detail in \citet{Garofali2018}. With precise astrometric alignment between the {\it HST} and {\it Chandra} imaging, we are able to search for {\it HST} counterparts to N604-WRX within the {\it Chandra} positional uncertainty ($<$ 0\farcs7 post-alignment). We identified three possible counterparts in {\it HST} to N604-WRX; however, two of these were saturated in the archival {\it HST} WFPC2 imaging, and thus magnitudes were not recovered. The photometry and position are reported only for the faintest counterpart within the X-ray error circle of N604-WRX in \citet{Garofali2018}. 

The Local Group Galaxy Survey \citep[LGGS;][]{Massey2006} reports photometry for two sources within the X-ray positional uncertainty of N604-WRX: J013432.63+304704.5 with V magnitude = 17.39 and B-V=0.022, and J013432.64+304704.0 with V magnitude=17.19, and B-V=0.006. However, the LGGS photometry in the region is a blend of the three sources visible in the {\it HST} imaging, so these magnitudes are likely inflated relative to the intrinsic values for the three point sources. 

\begin{figure}
\centering
\includegraphics[width=0.5\textwidth,trim=350 0 350 0, clip]{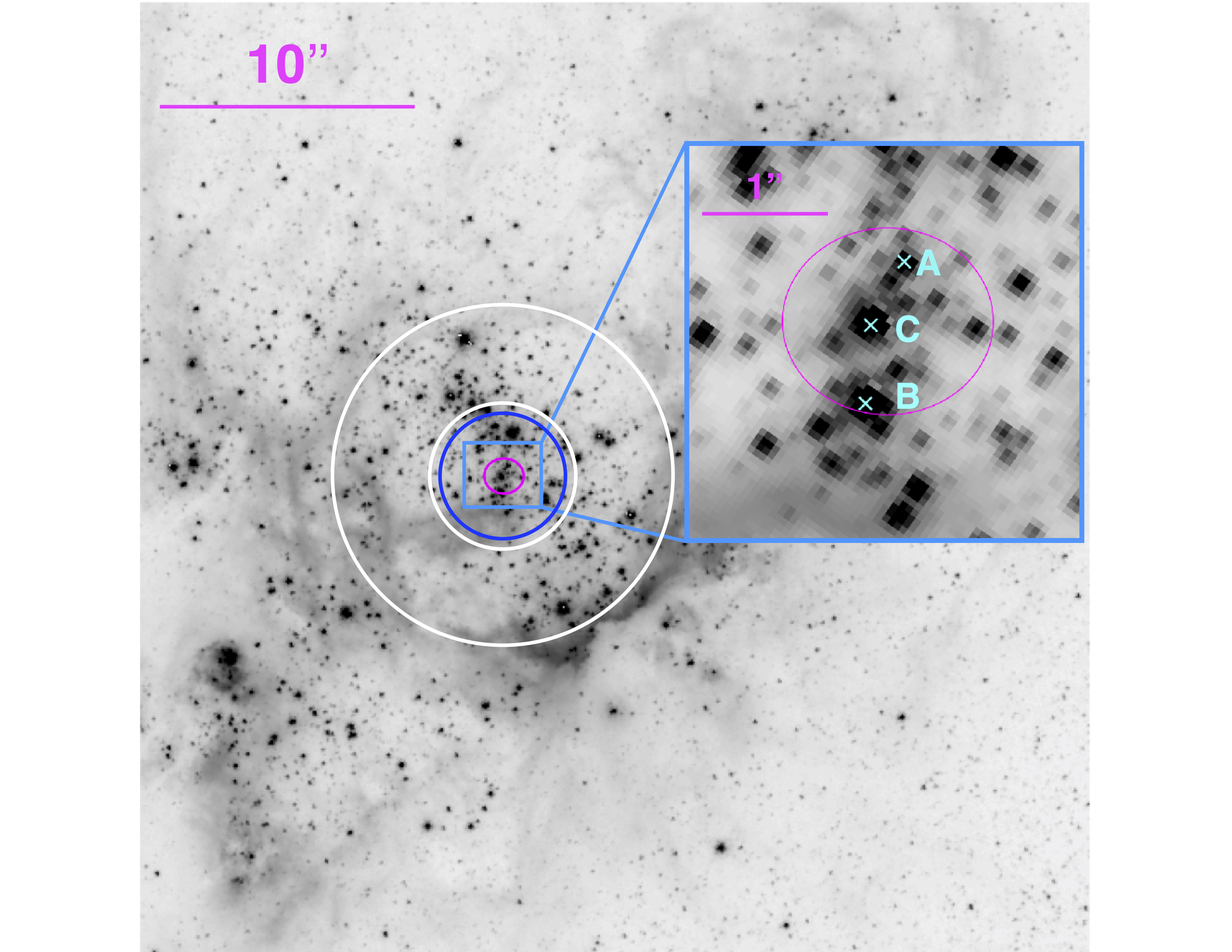}
\caption{{\it HST} ACS/WFC F475W image showing a portion of the giant H II region NGC 604 with the X-ray positional error for N604-WRX displayed in magenta. North is up and east is to the left. The X-ray spectral extraction regions ({\it dark blue}: source, {\it white}: background) are overlaid for reference. {\it Inset}: Zoom-in showing the three bright optical counterparts (A-C; labelled cyan crosses) within the X-ray positional uncertainty of N604-WRX.}\label{fig:hst_gem_targ}
\end{figure}

\begin{deluxetable*}{cccccccc}
 \tablecaption{ {\it HST} photometry of candidate optical counterparts to N604-WRX. All magnitudes are Vega magnitudes, and all positions are from {\it HST} and are aligned to 2MASS and the ChASeM33 survey \citep{Tullmann2011}. \label{tab:hstphot}}
\tablehead{
 \colhead{Identifier} & \colhead{RA} & \colhead{Dec} & \colhead{F275W}& \colhead{F336W} & \colhead{F475W} &\colhead{F814W}& \colhead{F160W}
 }

 \startdata 
N604-WRXa & 1:34:32.595 & +30:47:04.604 & 17.83 & 18.32 &  19.88 &  20.00 & 20.01 \\ 
N604-WRXb & 1:34:32.617 & +30:47:03.592 & 15.82 & 16.26 & 17.82 & 18.01 & 18.23 \\ 
N604-WRXc & 1:34:32.615 & +30:47:04.178 & 16.16 & 16.55 & 17.89 & 17.81 & 17.56 \\  
  \enddata
\end{deluxetable*}

To resolve the positions and magnitudes of all {\it three} possible counterparts within the X-ray error circle we therefore need refined {\it HST} imaging. To supplement the archival WFPC2 imaging, we made use of {\it HST} Advanced Camera for Surveys (ACS) imaging of the region from the forthcoming legacy survey of M33 (P.I. Dalcanton), which consists of imaging of 54 fields in M33 with the Wide Field Camera 3 (WFC3) in the UV (F275W, F336) and IR (F110W, F160W), and the Wide Field Camera (WFC) in the F475W and F814W filters. We align the ACS/WFC imaging to a common frame as the {\it Chandra} imaging, noting the same three stars from the archival WFPC2 imaging within the X-ray positional uncertainty of N604-WRX. 

The three possible counterparts to N604-WRX and their five filter photometry determined from the forthcoming {\it HST} legacy survey or M33 are listed in Table~\ref{tab:hstphot}. The stars are shown in an ACS/WFC F475W image of NGC 604 in Figure~\ref{fig:hst_gem_targ}, with the {\it Chandra} positional error (magenta), {\it Chandra} source extraction region (blue), and {\it Chandra} background extraction annulus (white) overlaid. We use the positions recovered from {\it HST} listed in Table~\ref{tab:hstphot} for targeted spectroscopic follow-up to determine the most likely counterpart to N604-WRX, as described in the next section. 

\subsection{Gemini Spectroscopy}\label{sec:gemreduce}

To determine the spectral type of the likely primary star in N604-WRX we obtained spectra of the two brightest stars within the X-ray error circle (targets N604-WRXb and N604-WRXc, Table~\ref{tab:hstphot}) using the Gemini Multiple-Object Spectrograph on Gemini-North (GMOS-N) \citep{Hook2004}. Neither N604-WRXb, nor N604-WRXc have clear previous spectral classifications in the literature (see Section~\ref{sec:results_spectype}). Target N604-WRXa was excluded from the proposed observations given its fainter magnitude compared to N604-WRXb/c and due to its small angular separation from star N604-WRXc ($<$ 0\farcs5), which posed an observational challenge for GMOS-N even under the best seeing conditions. 

To mitigate against crowding in NGC 604, both stars were observed using a blind-offset, with the slit placed at two different position angles on an offset bright star $\sim$ 1\farcm65 NE of NGC 604. The first target (N604-WRXb) was observed in July 2018, and the second target (N604-WRXc) was observed in August 2018, both in queue mode and at a mean air mass of 1.1. 

Spectra of both stars were taken using the 0\farcs25$\times$330" slit with the B600 grating centered first at 5200 $\AA$ and then at 5250 $\AA$ to compensate for gaps in the detectors. The observations consisted of two exposures per target, each 370 s. The image quality for both observations was 20th percentile, corresponding to a FWHM of 0\farcs5 for a point source in the R band observed at zenith\footnote{\url{http://www.gemini.edu/sciops/telescopes-and-sites/observing-condition-constraints\#ImageQuality}}. Standard baseline calibrations, including flat field images and CuAr lamp observations, were taken for flat fielding and wavelength calibrations, respectively. In addition, observations of the standard star Wolf 1346 were included for flux calibration. 

We reduced all data using the {\tt gemini} {\tt IRAF} package\footnote{\url{https://www.gemini.edu/node/11823}}. Bias and flat field images were created for each target using the tasks {\tt gbias} and {\tt gsflat}. The CuAr arc lamp observations and flux standard were reduced using the task {\tt gsreduce}, with the wavelength correction applied to the CuAr lamp using {\tt gswavelength}. The wavelength solution was applied to the flux standard and object using {\tt gstransform}, with sky subtraction applied to the flux standard only using {\tt gsskysub}. We did not perform automatic sky subtraction for either target due to issues with oversubtraction given the the confusing nebulosity from the surrounding H II region NGC 604. We performed additional sky subtraction for each target star by hand using the {\tt PyRAF} task {\tt splot} after the spectra were extracted. The spectra for both the standard and the target stars were extracted with {\tt gsextract}, with background subtraction also performed using this task. We derived the sensitivity function from the one-dimensional flux standard spectrum using {\tt gsstandard} and applied the function to both targets using {\tt gscalibrate}. We combined the spectra from the two exposures of each target using the {\tt PyRAF} task {\tt scombine} and an average of the pixels. 

\begin{figure}
\centering
\includegraphics[width=0.5\textwidth,trim=20 120 0 120, clip]{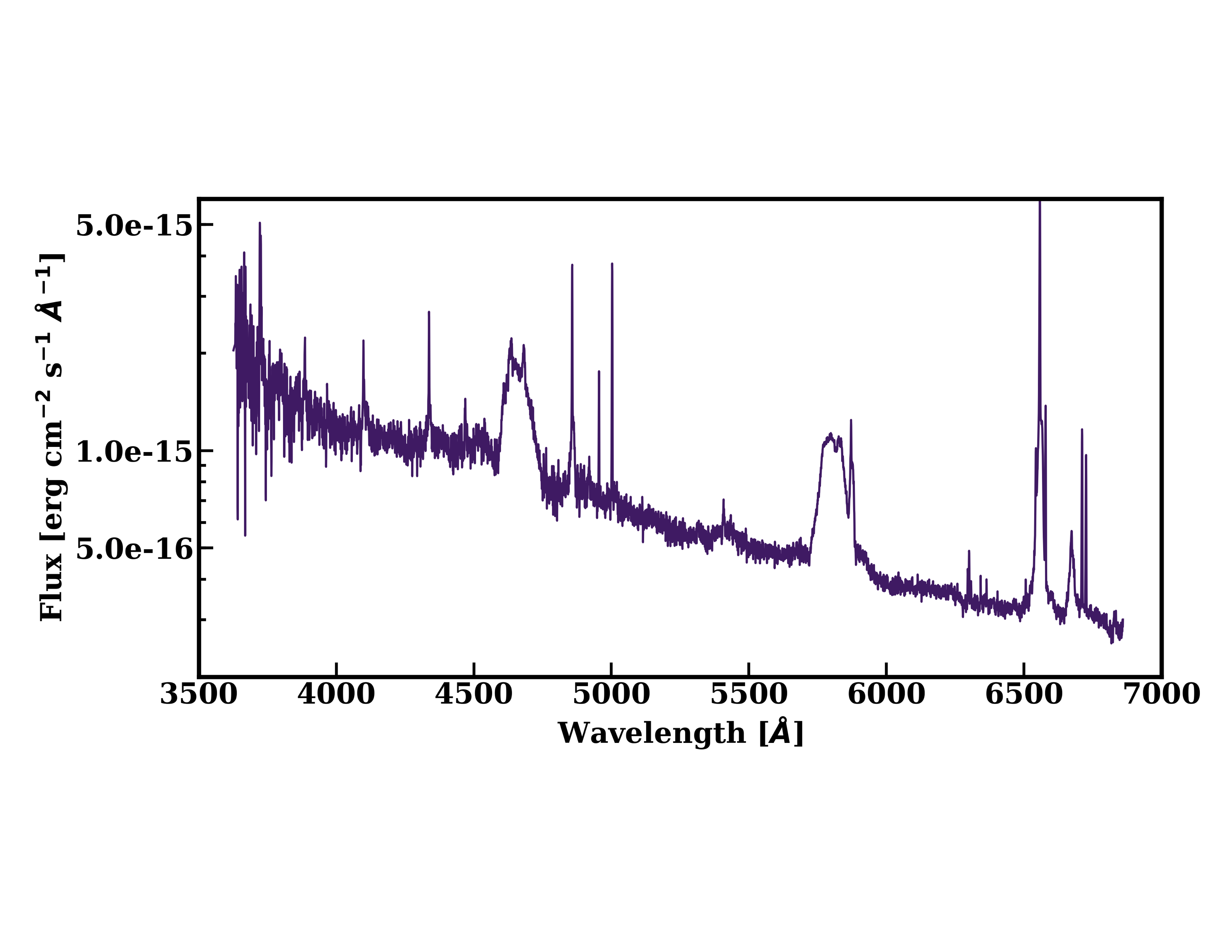}
\caption{Spectrum of target N604-WRXc, the likely primary for N604-WRX, taken with GMOS-N. The broad emission features in the spectrum are indicative of a WR star.}\label{fig:geminispec}
\end{figure}

The reduced spectra reveal strong, broad emission features in the spectrum of N604-WRXc, indicative of a WR star and therefore the most likely primary to N604-WRX. As N604-WRXc has no clear history of spectral classification in the literature prior to this work, we discuss our determination of spectral type for this star in Section~\ref{sec:results_spectype}. 

Given the abundance of massive stars the central cavity in NGC 604 where N604-WRX resides, we consider the probability that the newly identified WR star N604-WRXc is a chance coincidence with the X-ray error circle (r = 0\farcs7) of N604-WRX. We estimate a radius of $\sim$ 7" for the central cavity of NGC 604 from the ACS/WFC F745W image from {\it HST} and find eight WR stars catalogued in the literature within this radius \citep{Drissen2008,Neugent2011}, yielding a chance coincidence probability of $\sim$ 8\% for a WR star within the positional uncertainty of N604-WRX. If we consider the entirety of NGC 604 (r $\sim$ 25") and its WR population (14 stars), we get a $\sim$ 1\% chance coincidence probability of a WR star with the X-ray error circle of N604-WRX. This low probability of a chance coincidence coupled with the fact that CWBs with WR primaries are relatively prevalent in the Galaxy and MCs leads us to consider the newly identified WR star N604-WRXc as the most likely optical counterpart to N604-WRX, and therefore the most likely primary in the CWB. We show the extracted spectrum from GMOS-N for N604-WRXc in Figure~\ref{fig:geminispec}, and discuss its spectral properties in more detail in Sections~\ref{sec:results_spectype} and ~\ref{sec:discuss_opt_bin}.

\section{Results}\label{sec:results}

In this section, we present the observational results supporting classification of N604-WRX as a CWB. In particular, we present our X-ray spectral modeling and the resultant X-ray luminosity and long-term X-ray light curve, as well as the classification of the likely primary star as a WC4 star. 

\subsection{X-ray Spectral Modeling}\label{sec:xspec}

\begin{figure*}
\centering
\includegraphics[width=\textwidth,trim=0 120 0 120, clip]{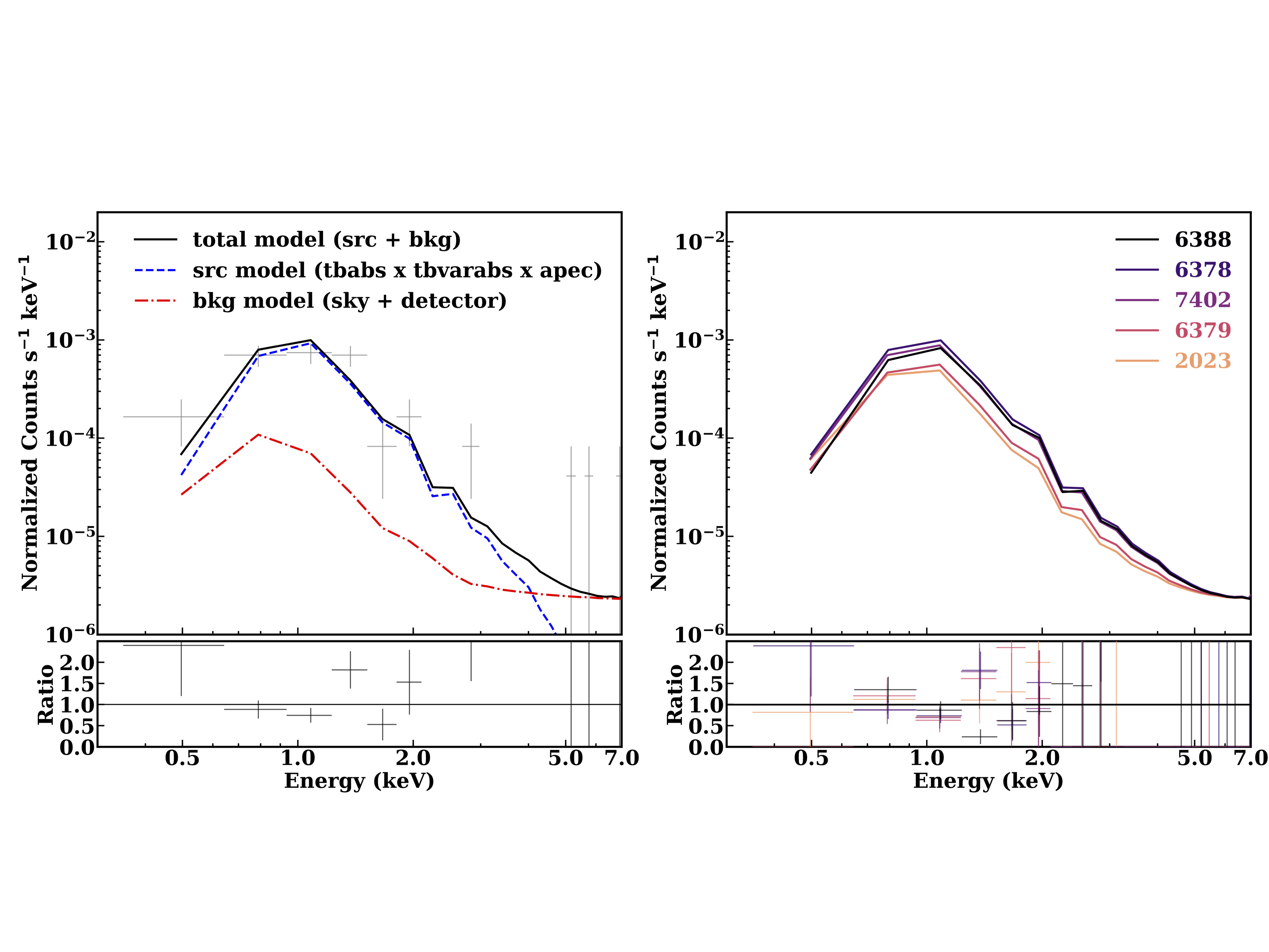}
\caption{{\it Left}: The best-fit total model (black solid line) for N604-WRX from {\tt XSPEC} for {\it Chandra} obs. ID 6378. The best-fit total model consists of both source (dashed blue) and background (dash-dot red) components. The spectrum has been rebinned to a minimum significance of five per bin for plotting purposes only. {\it Right}: Best-fit total (source + background) models for N604-WRX from all {\it Chandra} obs. IDs, where the spectra have again been rebinned for plotting purposes only.}\label{fig:xspec_fit}
\end{figure*}

We use {\tt XSPEC} v12.10.0c \citep{Arnaud1996} with the Cash statistic \citep{Cash1979} to simultaneously fit the ungrouped {\it Chandra} source and background spectra described in Section~\ref{sec:reduce}. We use the {\tt mcmc} routine in {\tt XSPEC} to estimate 90\% confidence intervals on the best-fit model parameters. Because N604-WRX is embedded in NGC 604, itself a source of soft X-ray emission, the spectral fitting process, and in particular the treatment of the background, requires customization and care beyond a simple background subtraction. In this section we describe our custom treatment of the background and our best-fit source plus background model, along with the inferred parameters for N604-WRX from the best-fit model. 

\subsubsection{Background Model: NGC 604}

Our background model is composed of two components: a sky and an instrumental component. The sky component is dominated by the soft emission from NGC 604, which we model following \citet{Tullmann2008} as a two component thermal plasma \citep[{\tt APEC};][]{Smith2001} with photoelectric absorption ({\tt phabs$\times$(apec + apec)}). We model the instrumental component as a power-law continuum coupled with Gaussian components to represent detector fluorescence lines. For the sky background, we fix $N_{\rm H}$ = 1.1 $\times$ 10$^{21}$ cm$^{-2}$ for the absorption in NGC 604 \citep{Leb2006} and scale the elemental abundances to Z = 0.65 $Z_{\odot}$ \citep{Vilchez1988,Magrini2007}.

We fit the sky and detector background model to {\it Chandra} spectra extracted from annular regions surrounding N604-WRX as described in Section~\ref{sec:chobs}. We allow the temperatures and normalizations of each background {\tt APEC} component to vary, as well as an overall multiplicative constant for both the sky and instrument background models, which are scaled for each observation by background size. We find best-fit temperatures and normalizations for the two-component {\tt APEC} model that are consistent with the values from \citet{Tullmann2008} for their regions ``C1", ``B1", and ``B2"--the main cavity of NGC 604 roughly coincident with the region hosting N604-WRX. We fix the temperatures and normalizations of the background {\tt APEC} components in our subsequent source plus background fits, allowing only the overall multiplicative constant to vary in order to characterize the background contribution to the source spectra. 

\subsubsection{Simultaneous Source and Background Fit: N604-WRX + NGC 604}\label{sec:bestfit}

We next perform a simultaneous fit to the source (N604-WRX) and background (NGC 604) spectra using a model for an absorbed diffuse plasma in collisional ionization equilibrium for the source ({\tt tbabs $\times$ tbvarabs $\times$ apec}), and a fixed background component as described in the preceding section. The {\tt tbabs} component in our source model accounts for Milky Way absorption, and the {\tt tbvarabs} component accounts for the hydrogen absorption column intrinsic to NGC 604. The {\tt APEC} \citep{Smith2001} component of the source model represents the X-ray emission from shocked material in the wind collision region of N604-WRX.

\begin{deluxetable*}{ccccc}
\tablewidth{\textwidth}
 \tablecaption{Absorbed and unabsorbed fluxes for N604-WRX from the best-fit {\tt tbabs $\times$ tbvarabs $\times$ apec} source model for the exposure with the highest signal-to-noise in each observation. Column (1): observation ID, column (2): net counts in the spectrum, column (3): absorbed and unabsorbed fluxes and associated errors in the 0.5--2.0 keV band from the best-fit model, column (4): absorbed and unabsorbed fluxes and associated errors in the 0.5--8.0 keV band from the best-fit model, and column (5): absorbed and unabsorbed luminosity and associated errors in the 0.5--2.0 keV band using the fluxes from column (3) and the assuming a distance of 817 kpc to M33 \citep{Freedman2001}. All errors are 90\% confidence intervals. All flux values from {\it XMM-Newton} exposures are upper limits.\label{tab:xspec_flux}}
\tablehead{
 \colhead{Obs. ID} & \colhead{Net Cts} & \colhead{$F_{\rm X,abs (unabs)}$ (0.5--2.0 keV)} & \colhead{$F_{\rm X,abs (unabs)}$ (0.5--8.0 keV)} & \colhead{$L_{\rm X,abs(unabs)}$ (0.5--2.0 keV)} \\ 
 \colhead{} & \colhead{} & \colhead{(10$^{-14}$ erg s$^{-1}$ cm$^{-2}$)} & \colhead{(10$^{-14}$ erg s$^{-1}$ cm$^{-2}$)}  & \colhead{(10$^{35}$ erg s$^{-1}$)}} 

\startdata 
2023 & 31 & 0.15 (0.16)$^{+0.03}_{-0.04}$ & 0.17 (0.18)$^{+0.03}_{-0.03}$ & 1.20 (1.28)$^{+0.24}_{-0.32}$ \\
6378 & 68  & 0.40 (0.41)$^{+0.06}_{-0.05}$ & 0.45 (0.46)$^{+0.11}_{-0.07}$ &  3.19 (3.27)$^{+0.48}_{-0.40}$ \\ 
6388 & 58 & 0.65 (0.68)$^{+0.26}_{-0.24}$ & 0.73 (0.76)$^{+0.25}_{-0.25}$ & 5.19 (5.43)$^{+2.08}_{-1.92}$  \\
6379 & 23  & 0.21 (0.22)$^{+0.07}_{-0.09}$ & 0.24 (0.25)$^{+0.06}_{-0.08}$ &1.68 (1.76)$^{+0.56}_{-0.72}$  \\ 
7402 &  27 & 0.35 (0.36) $^{+0.05}_{-0.05}$ & 0.40 (0.41)$^{+0.10}_{-0.06}$ & 2.80 (2.88)$^{+0.40}_{-0.40}$  \\
0650510101 (PN) & 511 & $<$ 0.40 (0.49) & $<$0.45 (0.52) & $<$ 3.19 (3.91)  \\
0672190301  (PN) & 440 & $<$ 0.22 (0.29) & $<$ 0.24 (0.30) &$<$ 1.76 (2.32) \\ 
0800350301 (MOS2) & 42 & $<$ 0.40 (0.72) & $<$ 0.44 (0.74) & $<$  3.19 (5.75)\\
\enddata 
\end{deluxetable*}

Although the integration time from {\it Chandra} is long ($\sim$ 320 ks) N604-WRX still has relatively few counts across all five observations ($\sim$ 200 counts), and we therefore cannot discriminate between more complex (i.e. two-temperature) models. Given the relatively low number of counts, we also fix the absorption in our model for both the line-of-sight absorption ({\tt tbabs}) to $N_{\rm H}$ = 5 $\times$ 10$^{20}$ cm$^{-2}$, and the intrinsic hydrogen column density ({\tt tbvarabs}) to  $N_{\rm H}$ = 1 $\times$ 10$^{21}$ cm$^{-2}$ \citep{Churchwell1999,Leb2006}. For the intrinsic absorption component ({\tt tbvarabs}) we fix the abundances of He, N, O, S, and Ne following \citet{Diaz1987}, \citet{Vilchez1988}, and \citet{Crockett2006}, and scale all other abundances to 0.65 $Z_{\odot}$ \citep{Vilchez1988,Asplund2009}.  

The total model is thus comprised of fixed sky and detector background components, and a source component where the plasma temperature and normalization are allowed to vary. We also allow an overall multiplicative constant, scaled by extraction region size, to vary for the source, sky background, and detector background components. 

Our best-fit {\tt tbabs $\times$ tbvarabs $\times$ apec} model returns {\it kT} = 0.98$^{+0.26}_{-0.11}$ keV, and {\it K} = 0.23$^{+0.08}_{-0.05}$ $\times$ 10$^{-5}$ cm$^{-5}$ with a null hypothesis probability of 0.49, where K is the normalization (10$^{-14}$ / (4$\pi$ {\it d$^{2}$}) $\int$ $n_{\rm e}$ $n_{\rm H}$ d{\it V}, with $n_{\rm e}$ and $n_{\rm H}$ as the electron and hydrogen number densities in cm$^{-3}$). We show the best-fit model (black solid line) for N604-WRX from {\it Chandra} obs. ID 6378 in the left panel of Figure~\ref{fig:xspec_fit}, with the source and background model components shown as the dashed blue and dash-dot red line, respectively. The right panel of Figure~\ref{fig:xspec_fit} shows the best-fit total model for N604-WRX for all {\it Chandra} obs. IDs. The spectra have been rebinned to a signal-to-noise of five in each bin for plotting purposes.

\newpage \subsection{N604-WRX X-ray Luminosity and Light Curve}\label{sec:results_lc}

We use the best-fit spectral model from Section~\ref{sec:bestfit} to determine the flux of N604-WRX in each {\it Chandra} observation and the upper limits on the flux in the {\it XMM-Newton} exposures. From these fluxes, we construct a long-term light curve of N604-WRX to look for variability.

In Table~\ref{tab:xspec_flux} we list the absorbed and unabsorbed fluxes in the 0.5--2.0 keV and 0.5--8.0 keV bands for each {\it Chandra} observation and upper limits on the flux for the {\it XMM-Newton} exposure with the highest signal-to-noise in each observation derived from the best-fit source model. {\it XMM-Newton} values are upper limits on the flux as N604-WRX is not resolved from the soft emission of NGC 604 in these observations. We exclude {\it XMM-Newton} observation 0650510301 from table, as N604-WRX was $>$ 8 arcmin off-axis in that observation, leading to very low signal-to-noise and unreliable upper limits on the flux. 

Fluxes and upper limits are derived from the best-fit source model using the {\tt flux} command in {\tt XSPEC}, with 90\% confidence intervals determined from the {\tt XSPEC} MCMC chains. We calculate luminosities from the model derived fluxes assuming a distance to M33 of 817 kpc \citep{Freedman2001}. We find a median unabsorbed flux across all {\it Chandra} observations of 3.6 $\times$ 10$^{-15}$ erg s$^{-1}$ cm$^{-2}$ ($L_{\rm X}$ =  2.9 $\times$ 10$^{35}$ erg s$^{-1}$) for N604-WRX. We discuss the implications for N604-WRX as a CWB based on the magnitude of its X-ray flux in Section~\ref{sec:discuss_xlum}.  

\begin{figure*}
\centering
\includegraphics[width=\textwidth,trim=0 120 0 120, clip]{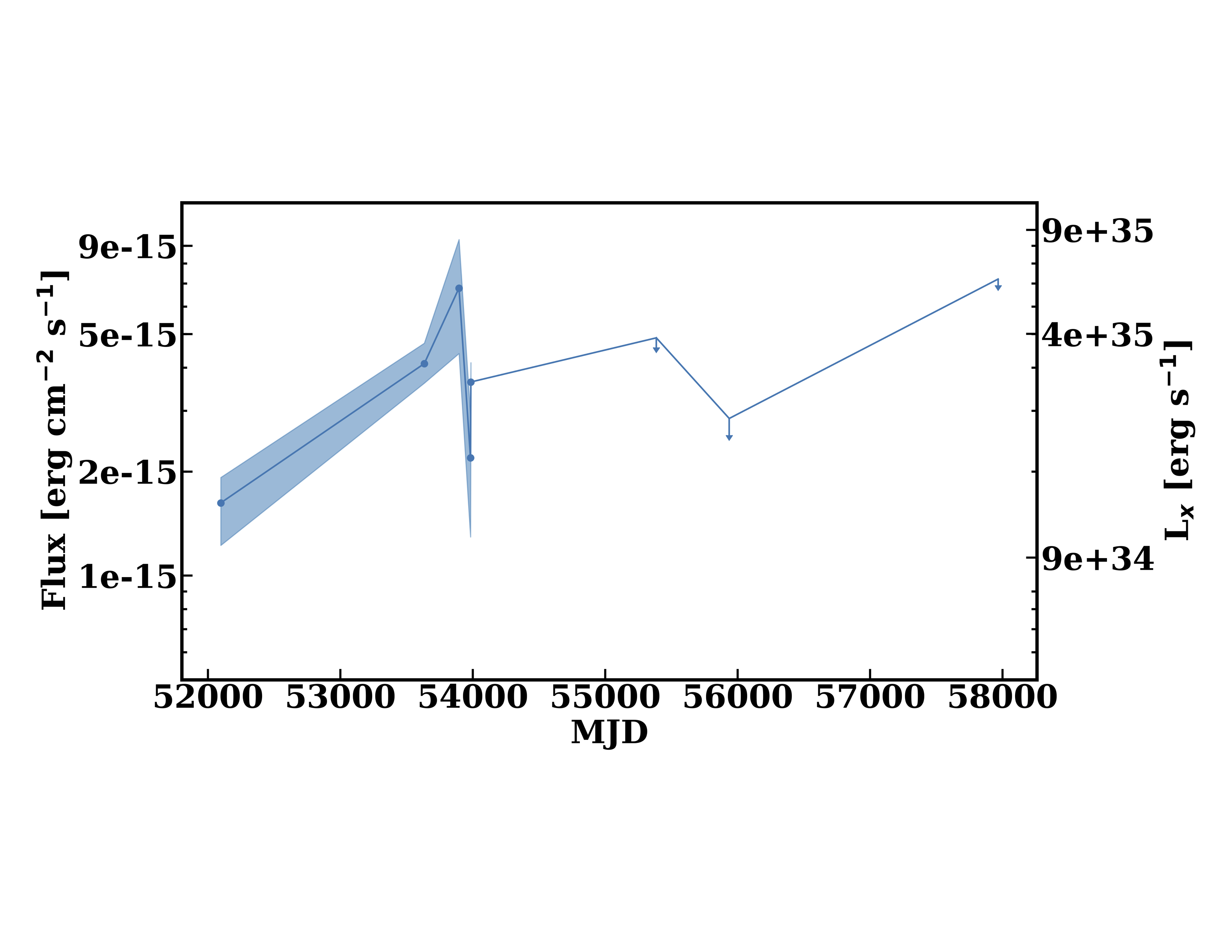}
\caption{Long-term X-ray light curve for N604-WRX (solid blue line). Fluxes are unabsorbed values in the 0.5--2.0 keV band taken from Table~\ref{tab:xspec_flux}. Blue shaded regions represent the 90\% confidence intervals on the fluxes. All luminosities are computed assuming a distance of 817 kpc to M33. Fluxes from the {\it XMM-Newton} observations with the highest signal-to-noise are shown as upper limits.}\label{fig:cwb_lc}
\end{figure*}

We construct a long-term light curve for N604-WRX, displayed in Figure~\ref{fig:cwb_lc}, from the unabsorbed 0.5--2.0 keV fluxes from Table~\ref{tab:xspec_flux}. The flux of N604-WRX varies by a factor of two to three between the {\it Chandra} observations. Given the low number of counts across all observations, we cannot reliably construct light curves in isolated energy bands, however we can look for evidence of any changes in the spectral hardness by computing the hardness ratio using counts in the soft (S; 0.5--2.0 keV) and hard (H; 2.0-8.0 keV) bands. We define the hardness ratio as $HR = (H-S)/(H+S)$, and compute this value and its associated 1$\sigma$ errors using the {\tt Bayesian Estimation of Hardness Ratios} \citep[BEHR;][]{BEHR} for all {\it Chandra} observations. Within the 1$\sigma$ errors on the hardness ratios we find no evidence for significant changes in the hardness of N604-WRX between {\it Chandra} observations. Without an orbital solution for this system we cannot further investigate possible phase-dependence in the spectral hardness or likewise produce a phase-resolved light curve for N604-WRX. In Section~\ref{sec:discuss_xvar} we discuss the variability and maximum X-ray luminosity reached by N604-WRX and their implications for an orbital period for N604-WRX.  

\subsection{Spectral Classification of the Candidate Primary Star}\label{sec:results_spectype}

Numerous past works have studied in detail the massive star population in M33, and specifically in NGC 604, leading to a large number of classifications for WR and OB stars in the region \citep[e.g.,][]{Conti1981,Massey1983,Hunter1996,Massey1996,Massey1998,Bruhweiler2003,Massey2006,Drissen2008,Neugent2011,Farina2012}. Below, we summarize sources in the literature that appear coincident with the position of N604-WRX, and then present our newly determined WC4 spectral classification for the likely primary star of N604-WRX.

The first identification of a WR star coincident with the position of N604-WRX comes from \citet{Drissen2008}. Based on their finder chart for NGC 604 \citep[Figure 3,][]{Drissen2008}, their star N604-WR3 appears to be coincident with the location of N604-WRX, and specifically N604-WRXc, the likely primary of N604-WRX. Unfortunately, tracing previous classifications in the literature from this identification is fraught due to issues with crowding. For example, \citet{Drissen2008} note possible previous classifications for N604-WR3 as WR star CM 12 \citep[][spectral type WNL]{Conti1981} and WR 136 \citep[][spectral type WNL]{Massey1998}. However, \citet{Drissen2008} also list their star N604-WR4 as the possible counterpart to CM 12/WR 136 (also listed as MC 75 \citep{Massey1983} and UIT 364 \citep{Massey1996}). We consider N604-WR4 as the more likely counterpart to these WR identifications in the literature based on our own visual inspection of the WR star positions, absolute magnitudes, and spectral classifications in NGC 604 from previous catalogs, as N604-WR4/CM 12/WR 136 are all consistent with a nitrogen-rich WR star that is brighter than, and to the north of N604-WRX and N604-WR3. 

Therefore, the only previous literature identification of a WR star coincident with the position of N604-WRX is N604-WR3 \citep[][spectral class WN]{Drissen2008}. Two sources from the LGGS \citep{Massey2006,Massey2016} also appear coincident with the position of N604-WRX, namely LGGS sources J013432.64+304704.0 (coincident with our Gemini target N604-WRXb), and J013432.63+304704.5 (coincident N604-WRXc, the most likely primary). However, neither LGGS target has a spectral classification from \citet{Massey2006}, and although \citet{Drissen2008} provide a WN classification for N604-WR3, they do not provide published spectroscopic confirmation. Given the lack of spectral data in the literature for the stars coincident with the position of N604-WRX, we turn to our GMOS-N spectra as the primary means of classifying the likely primary of N604-WRX.  

In Section~\ref{sec:gemreduce}, we determined GMOS-N target N604-WRXc to be the most likely primary for N604-WRX given the presence of strong, broad emission lines in the spectrum, indicative of the strong, fast winds of a WR star. The spectrum of the likely primary (Figure~\ref{fig:geminispec}) displays clear \ion{C}{4} $\lambda\lambda$5801-12 emission, and \ion{C}{3}-{IV} $\lambda$4650/\ion{He}{2} $\lambda$4686 emission complex, signifying a WC star. In particular, the presence of strong \ion{C}{4} $\lambda$5808, and the absence of \ion{C}{3} $\lambda$5696 in the spectrum signify that the likely primary to N604-WRX is a WC4 star \citep{Torres1986,Crowther1998}.

Puzzlingly, the strong \ion{C}{4} $\lambda$5808 and lack of distinguishing nitrogen features in the GMOS-N spectrum are at odds with the WN spectral classification listed in \citet{Drissen2008} for their star N604-WR3 that appears coincident with our GMOS-N target N604-WRXc. Given the unambiguous presence of \ion{C}{4} $\lambda$5808 in our GMOS-N spectrum, which is at odds with the WN designation from \citet{Drissen2008}, it seems unlikely that our targets are the same source, despite apparent positional overlap. This may not be altogether surprising given that this area of NGC 604 suffers from issues with crowding, and while our GMOS-N spectra were taken with a 0\farcs25 slit to avoid contamination from neighboring sources, the \citet{Drissen2008} spectral identifications were made with a 1\farcs5 slit. 

Our WC4 spectral classification for N604-WRXc is corroborated by Hectospec data from 2007 taken at the location of N604-WRX as part of spectroscopic follow-up for the ChASeM33 survey \citep{Tullmann2011}. In particular, the Hectospec spectra, which were taken with the Hectospec fiber positioned to be coincident with the {\it Chandra} position for N604-WRX from ChASeM33, show the same broad carbon emission features as our GMOS-N spectrum (Pete Challis, private comm.). Based on this analysis, we therefore determine that the primary to N604-WRX is a WC4 star newly identified in this work. 

We attempt to model the spectrum of the WC4 primary using the CoMoving Frame GENeral (CMFGEN) stellar atmosphere and radiative transfer code \citep{Hillier1998}. We use the LMC star BAT99-11 as a starting template for our model, as it is WC4 star in a sub-solar metallicity environment similar to NGC 604 \citep{Vilchez1988,Magrini2007}. We find that we cannot simultaneously reproduce the luminosity of the observed WC4 primary star along with the emission line ratios and strengths using a standard WC4-like template. In particular, the observed WC4 primary is more luminous than a standard WC4 star, but the strengths of the emission lines in the observed spectrum are lower than would be expected if a single WC4 star such as BAT99-11 were simply scaled up in luminosity. Furthermore, scaling a standard WC4 spectral template to the luminosity of the observed WC4 primary is incapable of reproducing the emission line ratios in the observed spectrum. Given that a spectral model for a single WC4 star is a poor fit to the spectrum of the observed primary we consider it likely that another component is necessary to describe the spectrum. We discuss these properties of the WC4 primary and their implications for N604-WRX as a CWB in more detail in Section~\ref{sec:discuss_opt_bin}.

\section{Discussion}\label{sec:discuss}

In this section we discuss the evidence that N604-WRX is a CWB, namely the soft X-ray spectrum coupled with the strength of the X-ray luminosity, and the WC4 primary star spectrum, which is not well described as a single WC4 star. We further compare N604-WRX to other nearby, bright CWBs to place constraints the orbital period from the strength of its X-ray luminosity and light curve. 

\subsection{A Soft X-ray Source with a Wolf-Rayet Optical Counterpart}

The X-ray spectrum of N604-WRX coupled with the presence of a likely WR counterpart places important constraints on the origin of the X-ray emission. In particular, these properties are strongly indicative of a CWB origin, but inconsistent with N604-WRX as a high-mass X-ray binary (HMXB) or supernova remnant (SNR). 

The spectrum of N604-WRX is best described as a soft ($\sim$ 1 keV) thermal plasma. This spectral shape is much softer than the harder disk blackbody or power law spectral shape associated with HMXBs \citep{Done2007}. In fact, N604-WRX has very few counts at energies $>$ 2 keV in these obesrvations, which is inconsistent even with XRBs observed in the soft state \citep{Done2007}. On the other hand, the soft spectrum of N604-WRX is entirely consistent with observations of other extragalactic CWBs \citep[e.g.,][]{Guerrero2008}, and the $\sim$ 1 keV temperature is consistent with theoretical predictions for plasma temperatures in the wind collision region of CWBs \citep{Pri1976,Cher1976}. 

Beyond spectral shape, further inconsistencies with an HMXB classification arise when we consider the X-ray luminosity of N604-WRX, and that the likely optical counterpart is a WR star. Of the handful of HMXBs known to host WR star donors (WR-XRBs), all systems have high X-ray luminosities ($L_{x}$ $\sim$ 10$^{38}$ erg s$^{-1}$), short orbital periods ($\sim$ days), and are observed to be time variable in X-rays \citep{vanK1996,Prestwich2007,Silverman2008,Crowther2010,Liu2013,Maccarone2014,vdh2019}. The high X-ray luminosities achieved in WR-XRBs are assumed to be the result of accretion of strong, fast WR winds onto a BH \citep{vdh2017}. By contrast, the median X-ray luminosity of N604-WRX from our best-fit spectral model is $L_{x}$ $\sim$ 10$^{35}$ erg s$^{-1}$, nearly three orders of magnitude below what is expected if the X-ray luminosity is powered by accretion of WR star winds onto a $>$ 10 $M_{\odot}$ BH \citep[see][]{White1986}. 

Accretion of WR winds onto a NS could produce X-ray luminosities in the range observed for N604-WRX, albeit with harder spectra, and dependent on the binary separation \citep[see][]{White1986}. However, models of binary evolution suggest that WR-XRBs with NS compact objects rarely, if ever, survive as their high mass ratios result in coalescence during common envelope evolution \citep{vdh2017}. We therefore find it highly unlikely that N604-WRX is an HMXB, and specifically a WR-XRB, given its soft spectrum and WR counterpart.

The soft spectrum and association with a very young stellar population do not immediately preclude classification as an SNR, as SNRs are soft X-ray sources and, if they have core-collapse progenitors, are often found in the vicinity of massive stars \citep[e.g.,][]{Jennings2014}. However, past SNR surveys in M33 and studies of the cluster environment where N604-WRX is located indicate that it is unlikely to be an SNR. 

To begin with, N604-WRX does not appear in the SNR catalogs of \citet{Long2010}, \citet{LeeLee2014}, or \citet{Garofali2017}, indicating that it was not classified as an SNR on the basis of its narrow-band optical emission or X-ray hardness ratios. Furthermore, the stellar cluster in the western portion of NGC 604, of which N604-WRX is a member, has an extremely young inferred age of $\sim$ 3 Myr based on its substantial WR star population \citep{Hunter1996,Bruhweiler2003,MA2004}. Models suggest that energy input from supernovae would not make a substantial contribution $<$ 3 Myr post-starburst \citep[e.g.,][]{Leitherer1992}, and thus we expect few, if any, core-collapse supernovae to have gone off in this particular region since the initial starburst 3 Myr ago. Finally, previous studies of wind-blown bubbles in the western portion of NGC 604, where N604-WRX is located, find that the observed X-ray emission does not require any extra heating from supernovae or SNRs \citep[e.g.,][]{Yang1996,Tullmann2008}, and is in fact consistent with the X-ray luminosity expected from a wind-blown cavity emanating from extensive mass loss from the numerous OB and WR stars in the region \citep[e.g.,][]{Hunter1996,Bruhweiler2003,Drissen2008}.

We therefore find the observed X-ray properties of N604-WRX to be inconsistent with classification as more commonly observed extragalactic X-ray sources, namely HMXBs or SNRs, and instead find the spectrum and X-ray luminosity of N604-WRX to be most consistent with a CWB. 

\subsection{X-ray Luminosity: Brighter than a Single Wolf-Rayet Star or Cluster}\label{sec:discuss_xlum}

\begin{figure*}
\centering
\includegraphics[width=\textwidth,trim=0 190 0 190, clip]{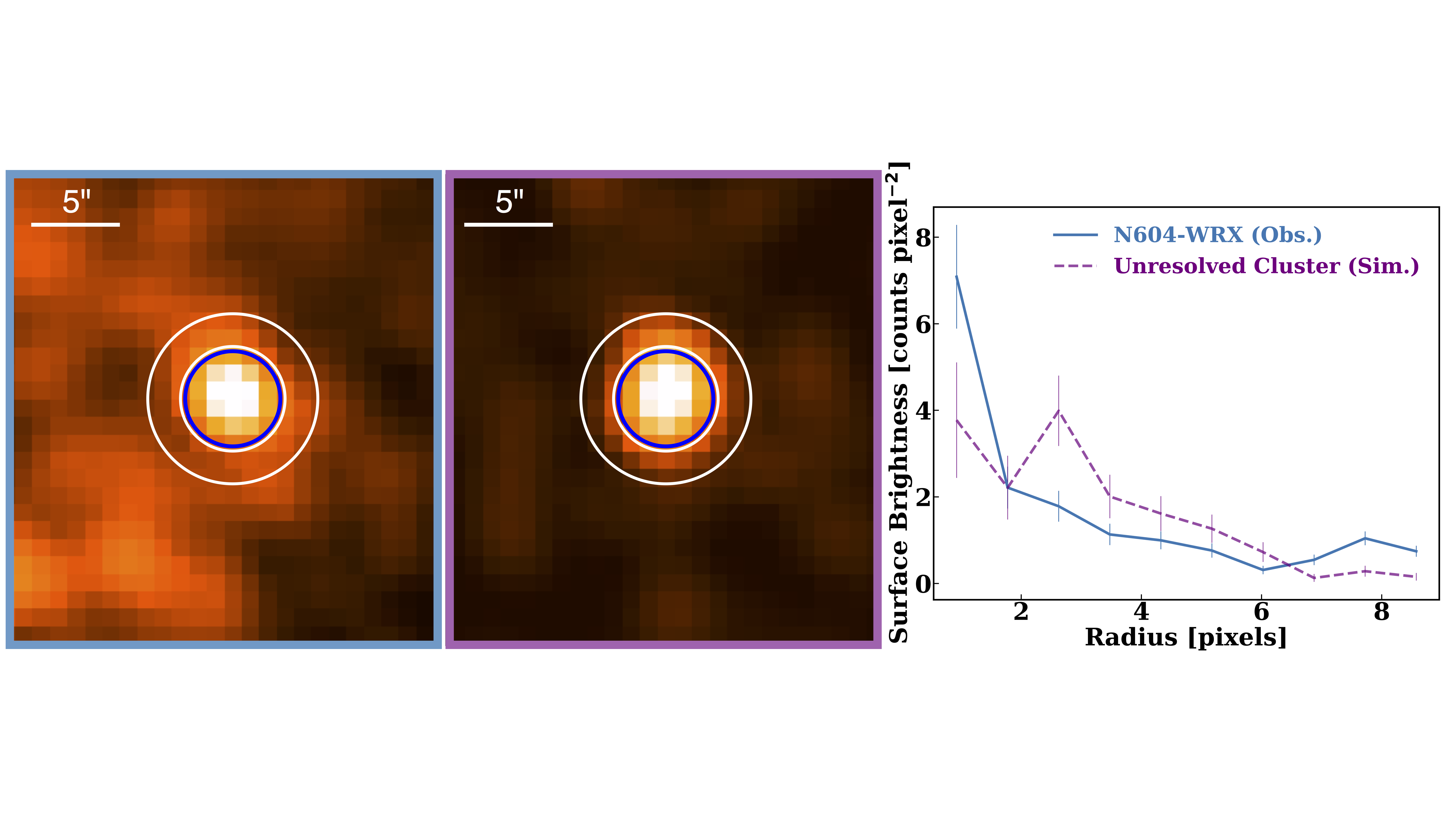}
\caption{{\it Left}: {\it Chandra} 0.2-8.0 keV mosaic of N604-WRX with source (blue) and background (white) extraction regions overlaid. Image has been binned into 1" pixels, and smoothed using a Gaussian filter with $\sigma$=3 {\it Center}: Simulated 0.2-8.0 keV {\it Chandra} image of an unresolved cluster of massive massive stars located within the source extraction region for N604-WRX, and at the location of N604-WRX on the detector. The image has been normalized by a factor of $>$ 300 compared to the left panel. The same source and background extraction regions as the left panel are overlaid for reference. The image has also been binned to 1" pixels and smoothed with a Gaussian filter with $\sigma$=3 for comparison with the {\it Chandra} observations. {\it Right}: The radial surface brightness profiles for N604-WRX (solid blue, left panel), and the simulated unresolved cluster of massive stars (dashed purple, center panel). The normalization mismatch between N604-WRX and the unresolved cluster, and extended nature of the cluster compared to N604-WRX as shown here rules out an unresolved cluster of massive stars as the source of X-ray emission for N604-WRX.}\label{fig:marxsim}
\end{figure*}

The X-ray luminosity inferred from our best-fit spectral model ($L_{\rm X}$ $\sim$ 10$^{35}$ erg s$^{-1}$) likewise places constraints on the nature of N604-WRX. As discussed in the previous section, this luminosity is much fainter than the luminosities observed for WR-XRBs. However, {\it single} massive stars also exhibit low luminosity, soft X-ray emission, so the possibility remains that N604-WRX is a single X-ray emitting massive star or an unresolved collection of X-ray emitting stars in a cluster. We discuss and dispel both possibilities below in favor of N604-WRX as a CWB. 

Theoretical predictions and empirical measurements of the X-ray emission emanating from embedded shocks in the winds of massive stars predict X-ray luminosities of $L_{\rm X}$ $\lesssim$ 10$^{33}$ erg s$^{-1}$ \citep[e.g.,][]{Harnden1979,Cass1981,Chle1989}. This value is well below the limiting luminosity of recent deep X-ray surveys of M33 \citep{Tullmann2011, Williams2015}, which alone makes detection of single X-ray emitting OB or WR stars in M33 unlikely unless such stars in M33 exceed theoretical luminosity predictions and empirical luminosity constraints by more than an order of magnitude. Considering that binaries, and CWBs in particular, are observed to be on average more X-ray luminous than single OB or WR stars \citep[e.g.,][]{Pollock1987,Chle1989}, the detection of N604-WRX in M33 is more plausibly explained by a binary origin, rather than a single WR star. 

To determine whether it is possible that N604-WRX is simply an X-ray emitting massive binary, as opposed to a CWB with a bright wind collision region, we turn to the canonical empirical $L_{\rm X}$ / $L_{\rm bol}$ $\sim$ 10$^{-7}$ scaling relation for X-ray emitting massive stars \citep{Long1980,Cass1981,Seward1982,Scior1990,Chle1991,Berghofer1997,Sana2006,Naze2011}. Conformity to this empirical relation signifies that the dominant source of X-ray emission is shocks embedded in the winds of massive stars, and it is applicable to both single and binary massive stars. The only exceptions to the relation are X-ray bright CWBs, where the dominant source of X-ray emission is the macroscopic wind collision region, as opposed to the embedded shocks in the winds of the massive star components \citep[e.g.,][]{Rauw2002}. 

To test where N604-WRX lies relative to the $L_{\rm X}$ / $L_{\rm bol}$ relation we take $M_{\rm V}$ = -7.0 for the system (see discussion in Section~\ref{sec:discuss_opt_bin}) and, after applying a bolometric correction appropriate for a WC4 star \citep{Nugis2000}, derive $M_{\rm bol}$ = -12.58. From this rough estimate of the bolometric magnitude we derive $L_{\rm bol}$ $\sim$  3 $\times$ 10$^{40}$ erg s$^{-1}$, which corresponds to an $L_{\rm X}$ / $L_{\rm bol}$ $\sim$ 1$\times$ 10$^{-5}$, using the median unabsorbed X-ray flux in the 0.5--8.0 keV band from Table~\ref{tab:xspec_flux}. This value is about two dex greater than the canonical $L_{\rm X}$ / $L_{\rm bol}$ relation which has a dispersion of 0.21 dex \citep{Naze2011}, indicating that the X-ray emission from N604-WRX more likely originates from a bright wind collision region rather than shocks in the winds of a massive star or stars. 

Finally, we wish to test whether the X-ray luminosity of N604-WRX can instead be explained as a collection of unresolved X-ray emitting massive stars in a cluster. To test for this possibility we first tabulated the number of OB and WR stars in the source extraction region (dark blue region, Figure~\ref{fig:chandra_rgb}). We selected known WR stars from the catalogs of \citet{Drissen2008} and \citet{Neugent2011}, accounting for duplicates, and included our newly characterized WC4 star, though under the assumption it is not a CWB. We counted OB stars in the region from the catalogs of \citet{Hunter1996} and \citet{Massey2016}, again accounting for duplicates. We included all spectroscopic identifications from \citet{Massey2016} within our extraction region, and, following \citet{Hunter1996}, all stars in their catalog with F555W-F814W colors $<$ 0.3, and $M_{\rm F555W}$ $<$ -4 (assuming E(B-V) = 0.08), corresponding roughly to stars of spectral type O9.5 V and earlier. Finally, we cross matched these OB identifications from \citet{Hunter1996} and \citet{Massey2016} with massive stars from the {\it HST} photometry reduced in \citet{Garofali2018} to again check for duplicates and look for any stars previously missed due to crowding. In total, we identify 4 WR stars and 18 OB stars in the {\it Chandra} source extraction region.

We used the published magnitudes and colors for all 22 stars to estimate order-of-magnitude luminosities using the PARSEC-COLIBRI stellar isochrones \citep{Marigo2017}. We compared the stars' magnitudes and colors to $\sim$ 3 Myr isochrones \citep[roughly the age of the cluster in NCG 604,][]{Hunter1996,GD2000,MA2004,Eldridge2011} at Z = 0.65 $Z_{\odot}$ \citep{Vilchez1988,Magrini2007}, assuming a Kroupa initial mass function and reddening appropriate to the region \citep[e.g.,][]{Massey1995}. In this way, we derived approximate bolometric luminosities for each star in the extraction region which we then converted to X-ray luminosities assuming $L_{\rm X}$ $\sim$ 10$^{-7} L_{\rm bol}$, which is appropriate for both single X-ray emitting massive stars and binaries where the wind collision region does not dominate the X-ray luminosity \citep[e.g.,][]{Berghofer1997,Naze2011}. 

We next created simulated event lists for each star in the region assuming a point source, and using {\tt MARX} \citep{marx2017} with the following inputs: the photon fluxes inferred from our $L_{\rm X}$ measurements (assuming D = 817 kpc \citep{Freedman2001}), the known positions of the stars from {\it HST}, and a simple {\tt tbabs $\times$ apec} spectral model with $N_{\rm H}$ = 1.1 $\times$ 10$^{21}$ cm$^{-2}$ \citep{Leb2006}, kT= 1 keV, and Z = 0.65 $Z_{\odot}$ \citep{Vilchez1988,Magrini2007} produced using {\tt Sherpa} \citep{Freeman2001}. Additionally, we used {\tt MARX} to create an event list representative of the diffuse background emission in the western portion of NGC 604 using the spectral model and associated flux taken from \citet{Tullmann2008} for their region ``B1". 

For each star, as well as for the background, we set the exposure time to zero, and ``NumRays" to the number of counts we wished to detect. The desired background count level was taken from \citet{Tullmann2008} again for their region ``B1", and we set each of the 22 stars to detect four photons each, amounting to a total of $\sim$ 90 counts for all sources, similar to the total number of counts detected for N604-WRX in a single deep {\it Chandra} exposure. We merged the individual event lists for each star together with the diffuse background event list using {\tt dmmerge} and additionally imposed a 2 $\times$ 2 binning to create a final simulated image that we could compare to our {\it Chandra} observations of N604-WRX.

In the center panel of Figure~\ref{fig:marxsim} we display a simulated image of the X-ray emission from an unresolved cluster of massive stars within the source extraction region (blue) at the location of N604-WRX, while in the left panel we show a 0.2-8.0 keV mosaic of the {\it Chandra} observations of N604-WRX. The right panel of Figure~\ref{fig:marxsim} shows the radial surface brightness profiles for both N604-WRX (solid blue line), and the simulated unresolved cluster (dashed purple line). 

While both images contain of order the same number of counts within the source extraction region (blue circle), the {\it Chandra} mosaic represents $\sim$ 300 ks of observation time while the simulated image from {\tt MARX} requires $>$ 100 Ms to achieve even a modest number of counts in the source extraction region. This extremely long integration time is due to the faint X-ray luminosities of the individual massive stars in the region which, based on their inferred bolometric luminosities, all have $L_{\rm X}$ $\sim$ 10$^{31}$--10$^{32}$ erg s$^{-1}$. Furthermore, as shown in the right pane of Figure~\ref{fig:marxsim}, the simulated unresolved cluster is a slightly extended source of X-ray emission, while N604-WRX is a point source. Given the slightly extended nature of the emission from this simulated ensemble of massive stars and the extremely long integration time necessary to obtain such an image we rule out the possibility that the X-ray emission from N604-WRX emanates from an unresolved cluster of massive stars.  

\begin{figure*}
\centering
\includegraphics[width=\textwidth,trim=20 120 0 120, clip]{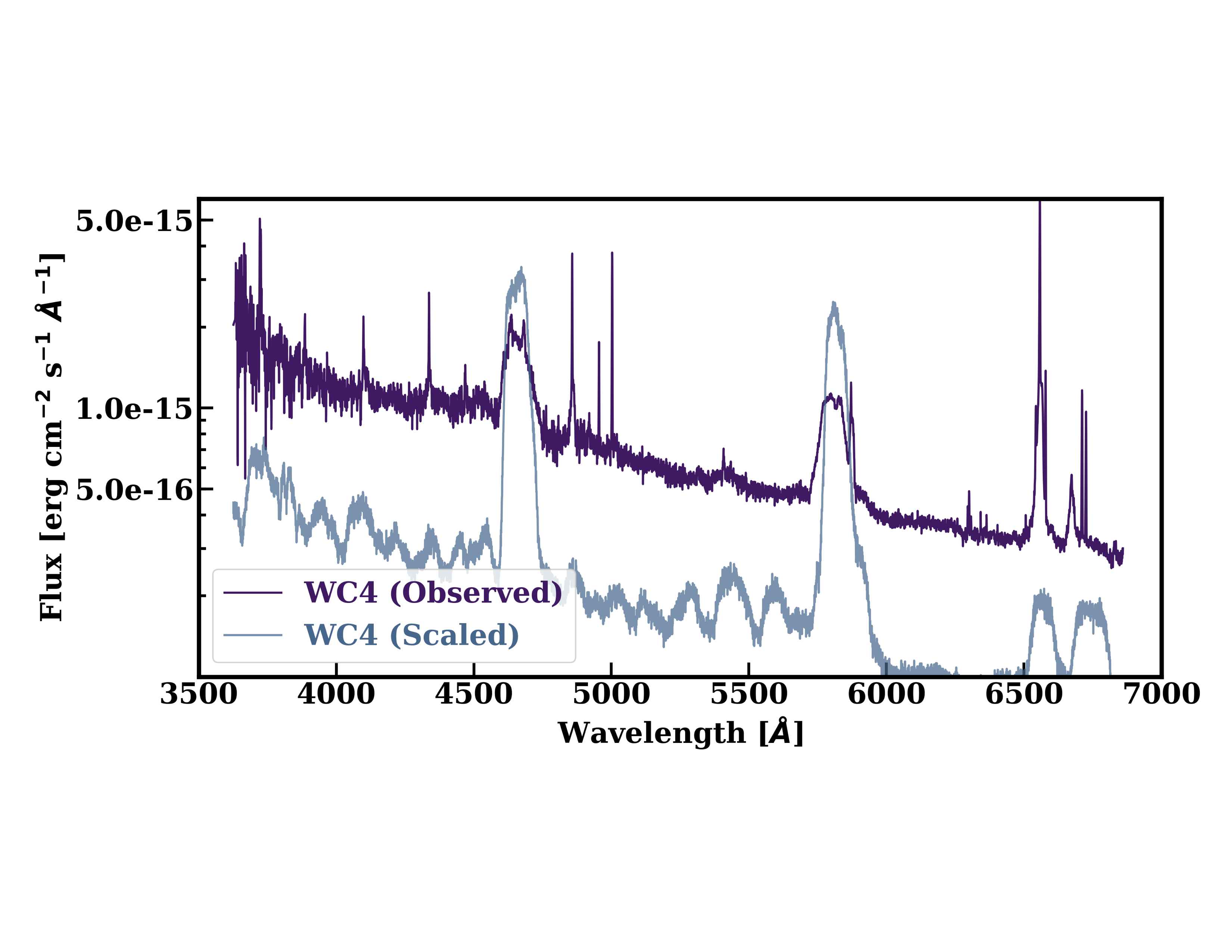}
\caption{Observed WC4 primary star spectrum from GMOS-N (purple), overlaid with the spectrum of an LMC WC4 star (blue) scaled to match the apparent magnitude of the observed spectrum. The spectrum of the likely WC4 primary to N604-WRX is not well-described by a WC4 star scaled up in luminosity.}\label{fig:simspec}
\end{figure*}

\subsection{Evidence for a Companion from the Primary Star Spectrum}\label{sec:discuss_opt_bin}

The WC4 spectral type of the likely primary coupled with its exceptional brightness in the optical and X-rays provide further evidence that N604-WRX is a CWB. Single WC stars are observed to be extremely faint sources in X-rays ($L_{x}$ $\sim$ 10$^{30}$ erg s$^{-1}$), owing to the high opacity of their metal-rich winds \citep{Oskinova2003,Rauw2015}. Thus the WC4 star observed to be coincident with relatively high X-ray luminosity makes it extremely unlikely that the X-ray emission observed from N604-WRX can be ascribed to a single star.  

Just as the high X-ray luminosity indicates a likely CWB, the magnitude of the WC4 counterpart is also indicative of binarity. The WC4 star has $m_{\rm F475W}$ = 17.89 from the {\it HST} photometry, which corresponds to $m_{\rm V}$ $\sim$ 18 as determined from the GMOS-N spectrum and {\tt pysynphot}. Assuming a modest reddening of E(B-V) = 0.13 \citep{Massey1995,Hunter1996} and a distance modulus of 24.6 this corresponds to $M_{\rm V}$ $\sim$ -7.0 for the WC4 primary. For comparison, WC stars in M33 typically have $M_{\rm V}$ = -3.0 -- -5.5 \citep[e.g.,][]{Massey1983}, while Galactic WC4 stars are observed to have average $M_{v}$ = -3.3 \citep[n.b., narrow band $v$;][]{Sander2012}. Thus, assuming even a moderate reddening, the likely primary to N604-WRX is $>$ 4$\times$ as bright as the brightest WC stars in M33, and $\sim$ 30$\times$ as bright as stars of the same spectral type in the Milky Way. 

The incredible brightness of the likely primary compared to stars of the same spectral type cannot be explained simply as a scaled up version of a WC4 star. In Figure~\ref{fig:simspec} we show the observed WC4 star spectrum from GMOS-N, as well as the spectrum of the LMC WC4 star BAT99-11 \citep[Brey 10, HD 32402;][]{Neugent2018} scaled to the apparent magnitude of the observed WC4 primary. Figure~\ref{fig:simspec} demonstrates that this simple approximation of a scaled-up WC star is not a good description of the observed WC4 primary in N604-WRX, as not only are the line strengths and relative line ratios incorrect, but the simulated continuum is also inconsistent with the observed WC4 star. 
 
The remarkably high $M_{\rm V}$ of the observed WC4 star relative to its spectral type may instead be explained by the presence of a companion. For example, if the companion is a late O supergiant with an absolute magnitude of $\sim$ -6.5 -- -7.0 \citep[e.g.,][]{Massey1998Mv} the observed apparent magnitude of the WC4 star as well as the relative weakness of its carbon emission features could be consistently explained. While a WC4 + O binary configuration would be entirely consistent with N604-WRX as a CWB, we see no clear evidence of an O companion (e.g., absorption lines) in the observed WC4 spectrum. 

To test whether we should expect to detect signatures of an O supergiant companion in the WC4 spectrum we create a composite WC + O spectrum in {\tt pysynphot} using scaled LMC spectra from an O7.5Iaf \citep[LMC173-2,][]{Massey2014} and WC4 star \citep[BAT99-11,][]{Neugent2018}. We scale the WC + O composite to match the apparent magnitude of the observed WC4 star, assuming the WC component contributes $\sim$ 40\% to the total luminosity. We then rebin the composite spectrum to GMOS-N resolution, and add noise consistent with the observed spectrum. 

\begin{figure*}
\centering
\includegraphics[width=\textwidth,trim=0 120 0 120, clip]{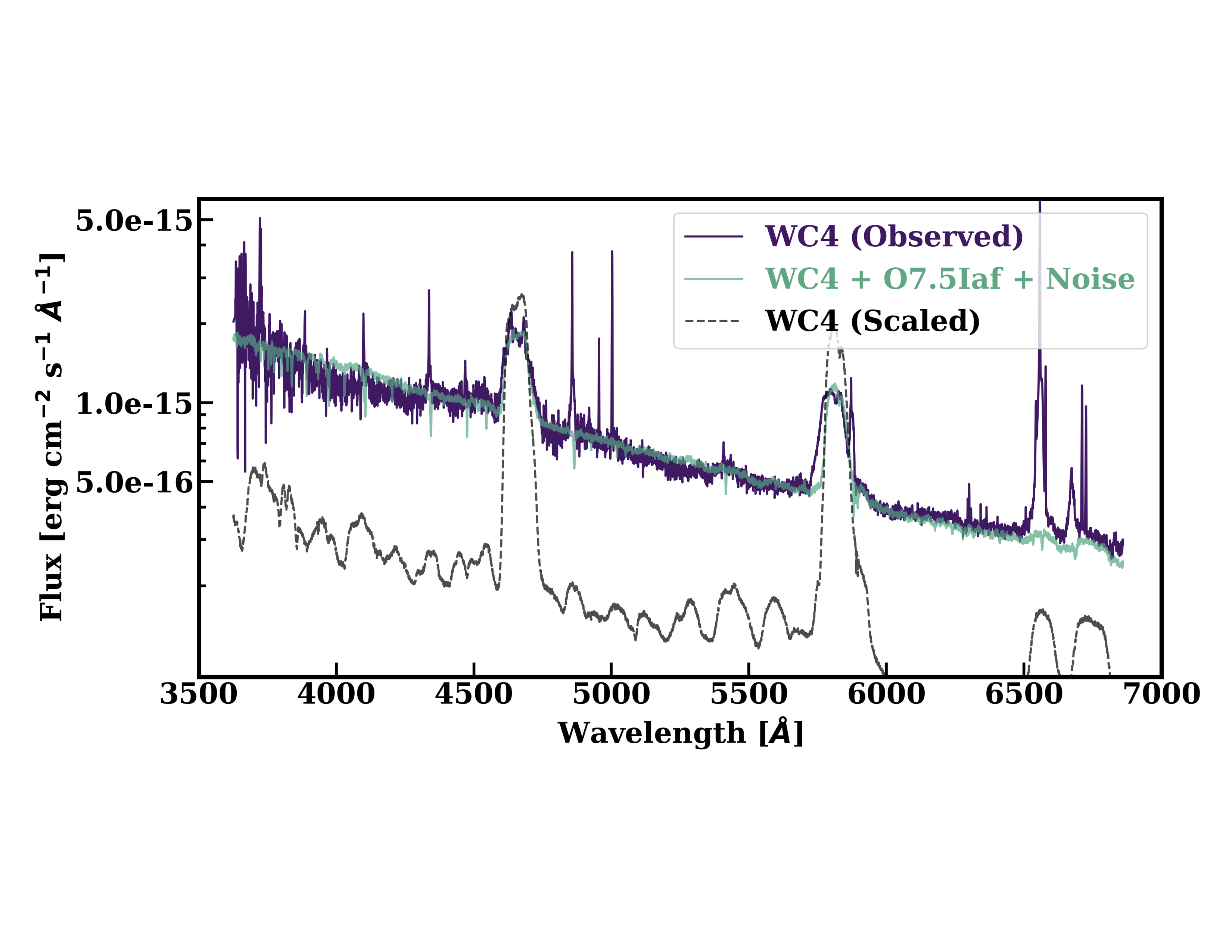}
\caption{Observed WC4 primary star spectrum from GMOS-N (purple) overlaid with a composite spectrum (green) constructed in {\tt pysynphot} from a combination of spectra from an LMC O7.5Iaf star and an LMC WC4 star. The composite spectrum has been scaled to match the magnitude of the observed WC4 star, rebinned to GMOS-N resolution, and has had noise added consistent with the observed spectrum. The O supergiant absorption lines are largely absent in the composite spectrum, while the strengths of the carbon emission features and continuum emission are consistent with the observed GMOS-N spectrum. For reference, we again show the spectrum of the LMC WC4 star BAT99-11 scaled to the luminosity of the observed WC4 star (black dashed line), which is a poor description of the spectrum of the observed WC4 primary.}\label{fig:simspec_ob}
\end{figure*}

The simulated WC + O composite spectrum is displayed in green in Figure~\ref{fig:simspec_ob}, and, notably, the absorption lines from the O supergiant component are largely erased with a realistic signal-to-noise and spectral resolution applied. Those absorption lines which are still visible in the composite are unfortunately coincident with strong nebular lines from NGC 604 in the observed spectrum. 

Importantly, the simulated WC + O spectrum shown in green in Figure~\ref{fig:simspec_ob} is able to match the continuum of the observed spectrum (purple) while preserving the relative strengths of the carbon emission features. This consistency between the observed spectrum and simulated composite indicates that, despite the lack of absorption line features in the GMOS-N spectrum, the observed WC4 spectrum is compatible with the presence of an O supergiant companion which contributes substantial flux. We discuss the prospects for follow-up observations to definitively determine the presence of a companion in the conclusions. 

\subsection{Comparison to Bright Colliding-Wind Binaries}\label{sec:discuss_xvar}

One of the most remarkable aspects of N604-WRX remains its high X-ray luminosity at maximum, which appears to exceed even that of the brightest CWBs in the Galaxy and MCs. In this section we discuss the X-ray luminosity and X-ray light curve of N604-WRX compared to other bright CWBs in order to place some constraints on the binary parameters giving rise to such an extreme X-ray luminosity. 

The unabsorbed X-ray luminosity of N604-WRX at maximum in the observed light curve is $L_{\rm X}$ $\sim$ 5.4 $\times$ 10$^{35}$ erg s$^{-1}$, a value which exceeds even the brightest CWBs in the LMC and Galaxy at their maxima: Mk 34 with $L_{\rm X}$ = 3.2 $\times$ 10$^{35}$ erg s$^{-1}$ in the LMC \citep{Pollock2018}, and $\eta$ Carinae with $L_{\rm X}$ = 2.4 $\times$ 10$^{35}$ erg s$^{-1}$ in the Galaxy \citep{Corcoran2017}. Both of these well-studied CWBs  are known to be quite massive, with recent spectroscopic measurements of Mk 34 indicating M $>$ 100 $M_{\odot}$ for each WN component of the binary \citep{Tehrani2019}, and a combined mass of  $>$ 100 $M_{\odot}$ for $\eta$ Car \citep{Hillier2001}. 

In terms of well-studied, bright CWBs with WC primaries there is WR 140, a WC7 + O4.5 system with an orbital period of $\sim$ 2900 days which reaches $L_{\rm X}$ = 4 $\times$ 10$^{34}$ erg s$^{-1}$ \citep{Marchenko2003,Pollock2005,Sugawara2015}, and Wr48a, a WC8 + WN8h binary with a $\sim$ 30 year period, which reaches as high as $L_{\rm X}$ = 3 $\times$ 10$^{35}$ erg s$^{-1}$ \citep{Zhekov2011,Zhekov2014WN8,Williams2012}. 

The common thread for these remarkably bright CWBs is, not surprisingly, very massive components with exceedingly strong winds. However, it does not appear that the spectral type of the primary uniquely determines which systems reach the highest X-ray luminosities, and the spectral type of the secondary is too often ill-constrained to make inferences as to its relative importance. Rather, it is likely a combination of the relative wind momenta of the binary components, as well as the orbital configuration (period, eccentricity, and viewer orientation) that set the observed X-ray luminosity. 

Of the systems discussed here, namely Mk 34, $\eta$ Car, WR 140, and WR48a, all are measured or presumed to have eccentric orbits and all are generally long period systems, with Mk 34 having the shortest orbital period at $\sim$ 155 days \citep{Pollock2018}, and all other systems having orbital periods $>$ 5 years \citep{Marchenko2003,Williams2012,Corcoran2017}. It therefore seems likely that N604-WRX, given its high X-ray luminosity, is likewise a long period system perhaps also with an eccentric orbit. 

Long period CWBs with eccentric orbits are more likely to approach the adiabatic limit in the wind collision region \citep{Pittard2002wind}, and in this case the star with the higher wind momentum (i.e., stronger wind) will largely set the X-ray luminosity. In such long period systems the wind components are also able to reach terminal velocity in the wind collision region, so variations in the X-ray luminosity will come about due to the changing separation of the stars and, theoretically, should scale as 1/$D$, where $D$ is the separation between the stars \citep[e.g.,][]{Gudel2009}. 

In theory, the 1/$D$ scaling for long period systems means their light curves can be described by this relation, with the X-ray luminosity reaching maximum at periastron. However, this is not always observed to be exactly the case. For example, WR 140 rises in flux as the system approaches periastron, but drops to a minimum just before conjunction, when the O star passes behind the more opaque winds of the WC7 primary \citep{Corcoran2011}. Deviations from the theoretical 1/$D$ light curve behavior for long period systems are thus indicative of the effects of changing line-of-sight absorption, or possibly the wind collision region becoming radiative at periastron.

As we do not yet have a full orbital solution for N604-WRX we can only speculate regarding periodicity in the light curve shown in Figure~\ref{fig:cwb_lc}. However, if we assume N604-WRX is a long period, eccentric CWB in the adiabatic limit like other bright CWBs then the light curve maximum in Figure~\ref{fig:cwb_lc} may represent an orbital phase close to periastron, with the rapid decrease shortly thereafter representing a phase close to conjunction. If we assume the most recent flux upper limit measurement from {\it XMM-Newton} also represents a phase close to maximum, then N604-WRX would have a period $\sim$ 4000 days, or $\sim$ 11 years. As the upper limits from {\it XMM-Newton} are derived from unresolved observations of N604-WRX with high background, we caution that deriving a period from these values is very speculative. More time-resolved observations will be necessary to determine where true minimum in the light curve lies, and thus derive an orbital solution. 

The lack of obvious periodicity in the light curve for N604-WRX is not unexpected given our limited observations. Although some nearby CWBs have shown evidence for periodicity in their X-ray light curves \citep[e.g.,][]{Lomax2015,Pollock2018}, more often CWBs exhibit phase-dependent variability in X-rays \citep[e.g.,][]{Pollock2006,Gosset2009}, with binary orbital periods later verified via spectroscopic follow-up \citep[e.g.,][]{Rauw2004,Bonanos2004,Gamen2006,Tehrani2019}. For CWBs with WC primaries in particular, orbital periods may be determined via IR observations if there is episodic dust formation near periastron \citep[e.g.,][]{Williams2012}. Clearly, multi-wavelength follow-up of N604-WRX is necessary to further characterize the orbital configuration of this remarkable system. 

\section{Summary \& Conclusions}\label{sec:conclude}

We present the identification N604-WRX, the first candidate CWB in M33, and the first such system observed beyond the MCs. The candidate CWB is a soft ($\sim$ 1 keV) source embedded in a very young cluster in the giant H II region NGC 604. Based on spectral modeling of N604-WRX we derive a median unabsorbed luminosity of $L_{\rm X}$(0.5--2.0 keV) = 2.9 $\times$ 10$^{35}$ erg s$^{-1}$, making this candidate CWB as bright as the most luminous known CWBs in the Galaxy and MCs, including $\eta$ Car and Mk 34. From our spectral models and all available archival {\it Chandra} and {\it XMM-Newton} observations we derive an X-ray light curve, demonstrating that N604-WRX varies by a factor of a few between observations, and reaches an unabsorbed luminosity of $L_{\rm X}$(0.5--2.0 keV) = 5.4 $\times$ 10$^{35}$ erg s$^{-1}$ at maximum. 

We identify a WR star coincident with the X-ray emission from N604-WRX, and consider this star to be the likely primary in the CWB. Using GMOS-N we are able to characterize the likely primary, for the first time, as a WC4 star whose high luminosity signifies the likely presence of a companion star, possibly a late O supergiant. Thus, we find N604-WRX to best be described as a WC4 + (O) CWB system. 

Given the distance to N604-WRX in M33 long integration times ($\sim$ 100s ks) are needed with current X-ray facilities to achieve even a modest increase in the number of detected counts for N604-WRX over currently available observations in the {\it Chandra} and {\it XMM-Newton} archives. Next generation X-ray facilities will therefore be of great importance to further identification and characterization of even the brightest extragalactic CWBs. 

The status of N604-WRX as a CWB can be verified via multiwavelength follow-up. In particular, follow-up spectroscopy of the likely WC4 primary with higher signal-to-noise will be necessary to look for signatures of a companion in the spectrum, and further provide a spectral classification for the secondary. Given the crowded field in NGC 604 {\it HST} may be necessary for such follow-up characterization to avoid contamination from nearby stars.

In order to determine an orbital solution for N604-WRX, which would also help cement classification as a CWB, dedicated radial velocity follow-up may be required, though confusing contamination from nebular emission from NGC 604 may make such a task more difficult. As we determined the likely primary to be a WC star, it may also be possible to look in the IR for periodic dust formation at periastron in order to derive an orbital period. 

Multiwavelength confirmation of N604-WRX as a CWB will be an important step towards expanding extragalactic studies of bright CWBs, and leveraging such sources as probes of massive star winds and massive star evolution in binaries.   

\section*{Acknowledgements}

Support for this work was provided by {\it Chandra} Award Number AR8-19006X issued by the {\it Chandra} X-ray Observatory Center, which is operated by the Smithsonian Astrophysical Observatory and on behalf of the National Aeronautics and Space Administration under contract NAS8-03060. This work is based on observations obtained as part of program GN-2018A-FT-210 at the Gemini Observatory, which is operated by the Association of Universities for Research in Astronomy, Inc., under a cooperative agreement with the NSF on behalf of the Gemini partnership: the National Science Foundation (United States), National Research Council (Canada), CONICYT (Chile), Ministerio de Ciencia, Tecnolog\'{i}a e Innovaci\'{o}n Productiva (Argentina), Minist\'{e}rio da Ci\^{e}ncia, Tecnologia e Inova\c{c}\~{a}o (Brazil), and Korea Astronomy and Space Science Institute (Republic of Korea). This work made use of {\tt PyRAF} and the Gemini {\tt IRAF} package. {\tt PyRAF} is a product of the Space Telescope Science Institute, which is operated by AURA for NASA. We would like to thank the referee for helpful comments in improving the manuscript.  K.G. would also like to thank Bret Lehmer and Kathryn Neugent for helpful discussions during the preparation of this manuscript. 

{\it Facilities:} {\it HST} (WFPC2 and ACS/WFC), Gemini-North (GMOS-N), {\it Chandra} (ACIS), {\it XMM-Newton} (EPIC-MOS and EPIC-pn), 
\software{Astropy \citep{astropy2013,astropy2018},
Matplotlib \citep{matplotlib}, CIAO \citep{ciao}, DOLPHOT and HSTPHOT \citep{Dolphin2000}, XSPEC \citep{Arnaud1996}}, MARX \citep{marx2017}, SAOImage DS9 \citep{ds9}
%%%%%%%%%%%%%%%%%%%%%%%%%%%%%%%%%%%%%%%%%%%%%%%%%%



\bibliographystyle{aasjournal}

\begin{thebibliography}{}
\expandafter\ifx\csname natexlab\endcsname\relax\def\natexlab#1{#1}\fi
\providecommand{\url}[1]{\href{#1}{#1}}
\providecommand{\dodoi}[1]{doi:~\href{http://doi.org/#1}{\nolinkurl{#1}}}
\providecommand{\doeprint}[1]{\href{http://ascl.net/#1}{\nolinkurl{http://ascl.net/#1}}}
\providecommand{\doarXiv}[1]{\href{https://arxiv.org/abs/#1}{\nolinkurl{https://arxiv.org/abs/#1}}}

\bibitem[{{Arnaud}(1996)}]{Arnaud1996}
{Arnaud}, K.~A. 1996, in Astronomical Society of the Pacific Conference Series,
  Vol. 101, Astronomical Data Analysis Software and Systems V, ed. G.~H.
  {Jacoby} \& J.~{Barnes}, 17

\bibitem[{{Asplund} {et~al.}(2009){Asplund}, {Grevesse}, {Sauval}, \&
  {Scott}}]{Asplund2009}
{Asplund}, M., {Grevesse}, N., {Sauval}, A.~J., \& {Scott}, P. 2009, Annual
  Review of Astronomy and Astrophysics, 47, 481,
  \dodoi{10.1146/annurev.astro.46.060407.145222}

\bibitem[{{Astropy Collaboration} {et~al.}(2013){Astropy Collaboration},
  {Robitaille}, {Tollerud}, {Greenfield}, {Droettboom}, {Bray}, {Aldcroft},
  {Davis}, {Ginsburg}, {Price-Whelan}, {Kerzendorf}, {Conley}, {Crighton},
  {Barbary}, {Muna}, {Ferguson}, {Grollier}, {Parikh}, {Nair}, {Unther},
  {Deil}, {Woillez}, {Conseil}, {Kramer}, {Turner}, {Singer}, {Fox}, {Weaver},
  {Zabalza}, {Edwards}, {Azalee Bostroem}, {Burke}, {Casey}, {Crawford},
  {Dencheva}, {Ely}, {Jenness}, {Labrie}, {Lim}, {Pierfederici}, {Pontzen},
  {Ptak}, {Refsdal}, {Servillat}, \& {Streicher}}]{astropy2013}
{Astropy Collaboration}, {Robitaille}, T.~P., {Tollerud}, E.~J., {et~al.} 2013,
  \aap, 558, A33, \dodoi{10.1051/0004-6361/201322068}

\bibitem[{{Astropy Collaboration} {et~al.}(2018){Astropy Collaboration},
  {Price-Whelan}, {Sip{\H o}cz}, {G{\"u}nther}, {Lim}, {Crawford}, {Conseil},
  {Shupe}, {Craig}, {Dencheva}, {Ginsburg}, {VanderPlas}, {Bradley},
  {P{\'e}rez-Su{\'a}rez}, {de Val-Borro}, {Aldcroft}, {Cruz}, {Robitaille},
  {Tollerud}, {Ardelean}, {Babej}, {Bach}, {Bachetti}, {Bakanov}, {Bamford},
  {Barentsen}, {Barmby}, {Baumbach}, {Berry}, {Biscani}, {Boquien}, {Bostroem},
  {Bouma}, {Brammer}, {Bray}, {Breytenbach}, {Buddelmeijer}, {Burke},
  {Calderone}, {Cano Rodr{\'{\i}}guez}, {Cara}, {Cardoso}, {Cheedella},
  {Copin}, {Corrales}, {Crichton}, {D'Avella}, {Deil}, {Depagne}, {Dietrich},
  {Donath}, {Droettboom}, {Earl}, {Erben}, {Fabbro}, {Ferreira}, {Finethy},
  {Fox}, {Garrison}, {Gibbons}, {Goldstein}, {Gommers}, {Greco}, {Greenfield},
  {Groener}, {Grollier}, {Hagen}, {Hirst}, {Homeier}, {Horton}, {Hosseinzadeh},
  {Hu}, {Hunkeler}, {Ivezi{\'c}}, {Jain}, {Jenness}, {Kanarek}, {Kendrew},
  {Kern}, {Kerzendorf}, {Khvalko}, {King}, {Kirkby}, {Kulkarni}, {Kumar},
  {Lee}, {Lenz}, {Littlefair}, {Ma}, {Macleod}, {Mastropietro}, {McCully},
  {Montagnac}, {Morris}, {Mueller}, {Mumford}, {Muna}, {Murphy}, {Nelson},
  {Nguyen}, {Ninan}, {N{\"o}the}, {Ogaz}, {Oh}, {Parejko}, {Parley}, {Pascual},
  {Patil}, {Patil}, {Plunkett}, {Prochaska}, {Rastogi}, {Reddy Janga},
  {Sabater}, {Sakurikar}, {Seifert}, {Sherbert}, {Sherwood-Taylor}, {Shih},
  {Sick}, {Silbiger}, {Singanamalla}, {Singer}, {Sladen}, {Sooley},
  {Sornarajah}, {Streicher}, {Teuben}, {Thomas}, {Tremblay}, {Turner},
  {Terr{\'o}n}, {van Kerkwijk}, {de la Vega}, {Watkins}, {Weaver}, {Whitmore},
  {Woillez}, {Zabalza}, \& {Astropy Contributors}}]{astropy2018}
{Astropy Collaboration}, {Price-Whelan}, A.~M., {Sip{\H o}cz}, B.~M., {et~al.}
  2018, \aj, 156, 123, \dodoi{10.3847/1538-3881/aabc4f}

\bibitem[{{Berghoefer} {et~al.}(1997){Berghoefer}, {Schmitt}, {Danner}, \&
  {Cassinelli}}]{Berghofer1997}
{Berghoefer}, T.~W., {Schmitt}, J.~H.~M.~M., {Danner}, R., \& {Cassinelli},
  J.~P. 1997, \aap, 322, 167

\bibitem[{{Bonanos} {et~al.}(2004){Bonanos}, {Stanek}, {Udalski},
  {Wyrzykowski}, {{\.Z}ebru{\'n}}, {Kubiak}, {Szyma{\'n}ski}, {Szewczyk},
  {Pietrzy{\'n}ski}, \& {Soszy{\'n}ski}}]{Bonanos2004}
{Bonanos}, A.~Z., {Stanek}, K.~Z., {Udalski}, A., {et~al.} 2004, \apj, 611,
  L33, \dodoi{10.1086/423671}

\bibitem[{{Bouret} {et~al.}(2012){Bouret}, {Hillier}, {Lanz}, \&
  {Fullerton}}]{Bouret2012}
{Bouret}, J.-C., {Hillier}, D.~J., {Lanz}, T., \& {Fullerton}, A.~W. 2012,
  \aap, 544, A67, \dodoi{10.1051/0004-6361/201118594}

\bibitem[{{Bruhweiler} {et~al.}(2003){Bruhweiler}, {Miskey}, \& {Smith
  Neubig}}]{Bruhweiler2003}
{Bruhweiler}, F.~C., {Miskey}, C.~L., \& {Smith Neubig}, M. 2003, \aj, 125,
  3082, \dodoi{10.1086/374988}

\bibitem[{{Cash}(1979)}]{Cash1979}
{Cash}, W. 1979, \apj, 228, 939, \dodoi{10.1086/156922}

\bibitem[{{Cassinelli} {et~al.}(1981){Cassinelli}, {Waldron}, {Sanders},
  {Harnden}, {Rosner}, \& {Vaiana}}]{Cass1981}
{Cassinelli}, J.~P., {Waldron}, W.~L., {Sanders}, W.~T., {et~al.} 1981, \apj,
  250, 677, \dodoi{10.1086/159414}

\bibitem[{{Cherepashchuk}(1976)}]{Cher1976}
{Cherepashchuk}, A.~M. 1976, Soviet Astronomy Letters, 2, 138

\bibitem[{{Chlebowski} \& {Garmany}(1991)}]{Chle1991}
{Chlebowski}, T., \& {Garmany}, C.~D. 1991, \apj, 368, 241,
  \dodoi{10.1086/169687}

\bibitem[{{Chlebowski} {et~al.}(1989){Chlebowski}, {Harnden}, \&
  {Sciortino}}]{Chle1989}
{Chlebowski}, T., {Harnden}, F.~R., J., \& {Sciortino}, S. 1989, \apj, 341,
  427, \dodoi{10.1086/167506}

\bibitem[{{Churchwell} \& {Goss}(1999)}]{Churchwell1999}
{Churchwell}, E., \& {Goss}, W.~M. 1999, \apj, 514, 188, \dodoi{10.1086/306941}

\bibitem[{{Clark} {et~al.}(2008){Clark}, {Muno}, {Negueruela}, {Dougherty},
  {Crowther}, {Goodwin}, \& {de Grijs}}]{Clark2008}
{Clark}, J.~S., {Muno}, M.~P., {Negueruela}, I., {et~al.} 2008, \aap, 477, 147,
  \dodoi{10.1051/0004-6361:20077186}

\bibitem[{{Cohen} {et~al.}(2011){Cohen}, {Gagn{\'e}}, {Leutenegger},
  {MacArthur}, {Wollman}, {Sundqvist}, {Fullerton}, \& {Owocki}}]{Cohen2011}
{Cohen}, D.~H., {Gagn{\'e}}, M., {Leutenegger}, M.~A., {et~al.} 2011, \mnras,
  415, 3354, \dodoi{10.1111/j.1365-2966.2011.18952.x}

\bibitem[{{Cohen} {et~al.}(2014){Cohen}, {Wollman}, {Leutenegger}, {Sundqvist},
  {Fullerton}, {Zsarg{\'o}}, \& {Owocki}}]{Cohen2014}
{Cohen}, D.~H., {Wollman}, E.~E., {Leutenegger}, M.~A., {et~al.} 2014, \mnras,
  439, 908, \dodoi{10.1093/mnras/stu008}

\bibitem[{{Conti} \& {Massey}(1981)}]{Conti1981}
{Conti}, P.~S., \& {Massey}, P. 1981, \apj, 249, 471, \dodoi{10.1086/159307}

\bibitem[{{Corcoran} {et~al.}(2011){Corcoran}, {Pollock}, {Hamaguchi}, \&
  {Russell}}]{Corcoran2011}
{Corcoran}, M.~F., {Pollock}, A.~M.~T., {Hamaguchi}, K., \& {Russell}, C. 2011,
  arXiv e-prints, arXiv:1101.1422.
\newblock \doarXiv{1101.1422}

\bibitem[{{Corcoran} {et~al.}(2017){Corcoran}, {Liburd}, {Morris}, {Russell},
  {Hamaguchi}, {Gull}, {Madura}, {Teodoro}, {Moffat}, {Richardson}, {Hillier},
  {Damineli}, \& {Groh}}]{Corcoran2017}
{Corcoran}, M.~F., {Liburd}, J., {Morris}, D., {et~al.} 2017, \apj, 838, 45,
  \dodoi{10.3847/1538-4357/aa6347}

\bibitem[{{Crockett} {et~al.}(2006){Crockett}, {Garnett}, {Massey}, \&
  {Jacoby}}]{Crockett2006}
{Crockett}, N.~R., {Garnett}, D.~R., {Massey}, P., \& {Jacoby}, G. 2006, \apj,
  637, 741, \dodoi{10.1086/498424}

\bibitem[{{Crowther} {et~al.}(2010){Crowther}, {Barnard}, {Carpano}, {Clark},
  {Dhillon}, \& {Pollock}}]{Crowther2010}
{Crowther}, P.~A., {Barnard}, R., {Carpano}, S., {et~al.} 2010, \mnras, 403,
  L41, \dodoi{10.1111/j.1745-3933.2010.00811.x}

\bibitem[{{Crowther} {et~al.}(1998){Crowther}, {De Marco}, \&
  {Barlow}}]{Crowther1998}
{Crowther}, P.~A., {De Marco}, O., \& {Barlow}, M.~J. 1998, \mnras, 296, 367,
  \dodoi{10.1046/j.1365-8711.1998.01360.x}

\bibitem[{{Dalcanton} {et~al.}(2009){Dalcanton}, {Williams}, {Seth}, {Dolphin},
  {Holtzman}, {Rosema}, {Skillman}, {Cole}, {Girardi}, {Gogarten},
  {Karachentsev}, {Olsen}, {Weisz}, {Christensen}, {Freeman}, {Gilbert},
  {Gallart}, {Harris}, {Hodge}, {de Jong}, {Karachentseva}, {Mateo}, {Stetson},
  {Tavarez}, {Zaritsky}, {Governato}, \& {Quinn}}]{Dalcanton2009}
{Dalcanton}, J.~J., {Williams}, B.~F., {Seth}, A.~C., {et~al.} 2009, \apjs,
  183, 67, \dodoi{10.1088/0067-0049/183/1/67}

\bibitem[{{Dalcanton} {et~al.}(2012){Dalcanton}, {Williams}, {Lang}, {Lauer},
  {Kalirai}, {Seth}, {Dolphin}, {Rosenfield}, {Weisz}, {Bell}, {Bianchi},
  {Boyer}, {Caldwell}, {Dong}, {Dorman}, {Gilbert}, {Girardi}, {Gogarten},
  {Gordon}, {Guhathakurta}, {Hodge}, {Holtzman}, {Johnson}, {Larsen}, {Lewis},
  {Melbourne}, {Olsen}, {Rix}, {Rosema}, {Saha}, {Sarajedini}, {Skillman}, \&
  {Stanek}}]{Dalcanton2012}
{Dalcanton}, J.~J., {Williams}, B.~F., {Lang}, D., {et~al.} 2012, \apjs, 200,
  18, \dodoi{10.1088/0067-0049/200/2/18}

\bibitem[{{Diaz} {et~al.}(1987){Diaz}, {Terlevich}, {Pagel}, {Vilchez}, \&
  {Edmunds}}]{Diaz1987}
{Diaz}, A.~I., {Terlevich}, E., {Pagel}, B.~E.~J., {Vilchez}, J.~M., \&
  {Edmunds}, M.~G. 1987, \mnras, 226, 19, \dodoi{10.1093/mnras/226.1.19}

\bibitem[{{Dolphin}(2000)}]{Dolphin2000}
{Dolphin}, A.~E. 2000, \pasp, 112, 1383, \dodoi{10.1086/316630}

\bibitem[{{Done} {et~al.}(2007){Done}, {Gierli{\'n}ski}, \&
  {Kubota}}]{Done2007}
{Done}, C., {Gierli{\'n}ski}, M., \& {Kubota}, A. 2007, Astronomy and
  Astrophysics Review, 15, 1, \dodoi{10.1007/s00159-007-0006-1}

\bibitem[{{Drissen} {et~al.}(2008){Drissen}, {Crowther}, {{\'U}beda}, \&
  {Martin}}]{Drissen2008}
{Drissen}, L., {Crowther}, P.~A., {{\'U}beda}, L., \& {Martin}, P. 2008,
  \mnras, 389, 1033, \dodoi{10.1111/j.1365-2966.2008.13633.x}

\bibitem[{{Eldridge} \& {Rela{\~n}o}(2011)}]{Eldridge2011}
{Eldridge}, J.~J., \& {Rela{\~n}o}, M. 2011, \mnras, 411, 235,
  \dodoi{10.1111/j.1365-2966.2010.17676.x}

\bibitem[{{Fari{\~n}a} {et~al.}(2012){Fari{\~n}a}, {Bosch}, \&
  {Barb{\'a}}}]{Farina2012}
{Fari{\~n}a}, C., {Bosch}, G.~L., \& {Barb{\'a}}, R.~H. 2012, \aj, 143, 43,
  \dodoi{10.1088/0004-6256/143/2/43}

\bibitem[{{Feldmeier} {et~al.}(1997){Feldmeier}, {Puls}, \&
  {Pauldrach}}]{Feld1997}
{Feldmeier}, A., {Puls}, J., \& {Pauldrach}, A.~W.~A. 1997, \aap, 322, 878

\bibitem[{{Freedman} {et~al.}(2001){Freedman}, {Madore}, {Gibson}, {Ferrarese},
  {Kelson}, {Sakai}, {Mould}, {Kennicutt}, {Ford}, {Graham}, {Huchra},
  {Hughes}, {Illingworth}, {Macri}, \& {Stetson}}]{Freedman2001}
{Freedman}, W.~L., {Madore}, B.~F., {Gibson}, B.~K., {et~al.} 2001, \apj, 553,
  47, \dodoi{10.1086/320638}

\bibitem[{{Freeman} {et~al.}(2001){Freeman}, {Doe}, \&
  {Siemiginowska}}]{Freeman2001}
{Freeman}, P., {Doe}, S., \& {Siemiginowska}, A. 2001, in Society of
  Photo-Optical Instrumentation Engineers (SPIE) Conference Series, Vol. 4477,
  Astronomical Data Analysis, ed. J.-L. {Starck} \& F.~D. {Murtagh}, 76--87

\bibitem[{{Fruscione} {et~al.}(2006){Fruscione}, {McDowell}, {Allen},
  {Brickhouse}, {Burke}, {Davis}, {Durham}, {Elvis}, {Galle}, {Harris},
  {Huenemoerder}, {Houck}, {Ishibashi}, {Karovska}, {Nicastro}, {Noble},
  {Nowak}, {Primini}, {Siemiginowska}, {Smith}, \& {Wise}}]{ciao}
{Fruscione}, A., {McDowell}, J.~C., {Allen}, G.~E., {et~al.} 2006, in
  \procspie, Vol. 6270, Society of Photo-Optical Instrumentation Engineers
  (SPIE) Conference Series, 62701V

\bibitem[{{Fullerton} {et~al.}(2006){Fullerton}, {Massa}, \&
  {Prinja}}]{Fullerton2006}
{Fullerton}, A.~W., {Massa}, D.~L., \& {Prinja}, R.~K. 2006, \apj, 637, 1025,
  \dodoi{10.1086/498560}

\bibitem[{{Gamen} {et~al.}(2006){Gamen}, {Gosset}, {Morrell}, {Niemela},
  {Sana}, {Naz{\'e}}, {Rauw}, {Barb{\'a}}, \& {Solivella}}]{Gamen2006}
{Gamen}, R., {Gosset}, E., {Morrell}, N., {et~al.} 2006, \aap, 460, 777,
  \dodoi{10.1051/0004-6361:20065618}

\bibitem[{{Garofali} {et~al.}(2018){Garofali}, {Williams}, {Hillis}, {Gilbert},
  {Dolphin}, {Eracleous}, \& {Binder}}]{Garofali2018}
{Garofali}, K., {Williams}, B.~F., {Hillis}, T., {et~al.} 2018, \mnras, 479,
  3526, \dodoi{10.1093/mnras/sty1612}

\bibitem[{{Garofali} {et~al.}(2017){Garofali}, {Williams}, {Plucinsky},
  {Gaetz}, {Wold}, {Haberl}, {Long}, {Blair}, {Pannuti}, {Winkler}, \&
  {Gross}}]{Garofali2017}
{Garofali}, K., {Williams}, B.~F., {Plucinsky}, P.~P., {et~al.} 2017, \mnras,
  472, 308, \dodoi{10.1093/mnras/stx1905}

\bibitem[{{Gonz{\'a}lez Delgado} \& {P{\'e}rez}(2000)}]{GD2000}
{Gonz{\'a}lez Delgado}, R.~M., \& {P{\'e}rez}, E. 2000, \mnras, 317, 64,
  \dodoi{10.1046/j.1365-8711.2000.03545.x}

\bibitem[{{Gosset} {et~al.}(2009){Gosset}, {Naz{\'e}}, {Sana}, {Rauw}, \&
  {Vreux}}]{Gosset2009}
{Gosset}, E., {Naz{\'e}}, Y., {Sana}, H., {Rauw}, G., \& {Vreux}, J.~M. 2009,
  \aap, 508, 805, \dodoi{10.1051/0004-6361/20077981}

\bibitem[{{G{\"u}del} \& {Naz{\'e}}(2009)}]{Gudel2009}
{G{\"u}del}, M., \& {Naz{\'e}}, Y. 2009, Astronomy and Astrophysics Review, 17,
  309, \dodoi{10.1007/s00159-009-0022-4}

\bibitem[{{Guerrero} \& {Chu}(2008)}]{Guerrero2008}
{Guerrero}, M.~A., \& {Chu}, Y.-H. 2008, The Astrophysical Journal Supplement
  Series, 177, 216, \dodoi{10.1086/587059}

\bibitem[{{G{\"u}nther} {et~al.}(2017){G{\"u}nther}, {Frost}, \&
  {Theriault-Shay}}]{marx2017}
{G{\"u}nther}, H.~M., {Frost}, J., \& {Theriault-Shay}, A. 2017, \aj, 154, 243,
  \dodoi{10.3847/1538-3881/aa943b}

\bibitem[{{Harnden} {et~al.}(1979){Harnden}, {Branduardi}, {Elvis},
  {Gorenstein}, {Grindlay}, {Pye}, {Rosner}, {Topka}, \&
  {Vaiana}}]{Harnden1979}
{Harnden}, F.~R., J., {Branduardi}, G., {Elvis}, M., {et~al.} 1979, \apj, 234,
  L51, \dodoi{10.1086/183107}

\bibitem[{{Hillier} {et~al.}(2001){Hillier}, {Davidson}, {Ishibashi}, \&
  {Gull}}]{Hillier2001}
{Hillier}, D.~J., {Davidson}, K., {Ishibashi}, K., \& {Gull}, T. 2001, \apj,
  553, 837, \dodoi{10.1086/320948}

\bibitem[{{Hillier} \& {Miller}(1998)}]{Hillier1998}
{Hillier}, D.~J., \& {Miller}, D.~L. 1998, \apj, 496, 407,
  \dodoi{10.1086/305350}

\bibitem[{{Hook} {et~al.}(2004){Hook}, {J{\o}rgensen}, {Allington-Smith},
  {Davies}, {Metcalfe}, {Murowinski}, \& {Crampton}}]{Hook2004}
{Hook}, I.~M., {J{\o}rgensen}, I., {Allington-Smith}, J.~R., {et~al.} 2004,
  \pasp, 116, 425, \dodoi{10.1086/383624}

\bibitem[{{Hunter} {et~al.}(1996){Hunter}, {Baum}, {O'Neil}, \&
  {Lynds}}]{Hunter1996}
{Hunter}, D.~A., {Baum}, W.~A., {O'Neil}, Jr., E.~J., \& {Lynds}, R. 1996,
  \apj, 456, 174, \dodoi{10.1086/176638}

\bibitem[{Hunter(2007)}]{matplotlib}
Hunter, J.~D. 2007, Computing In Science \& Engineering, 9, 90,
  \dodoi{10.1109/MCSE.2007.55}

\bibitem[{{Jennings} {et~al.}(2014){Jennings}, {Williams}, {Murphy},
  {Dalcanton}, {Gilbert}, {Dolphin}, {Weisz}, \& {Fouesneau}}]{Jennings2014}
{Jennings}, Z.~G., {Williams}, B.~F., {Murphy}, J.~W., {et~al.} 2014, \apj,
  795, 170, \dodoi{10.1088/0004-637X/795/2/170}

\bibitem[{{Joye} \& {Mandel}(2003)}]{ds9}
{Joye}, W.~A., \& {Mandel}, E. 2003, in Astronomical Society of the Pacific
  Conference Series, Vol. 295, Astronomical Data Analysis Software and Systems
  XII, ed. H.~E. {Payne}, R.~I. {Jedrzejewski}, \& R.~N. {Hook}, 489

\bibitem[{{Kudritzki} \& {Puls}(2000)}]{Kud2000}
{Kudritzki}, R.-P., \& {Puls}, J. 2000, \araa, 38, 613,
  \dodoi{10.1146/annurev.astro.38.1.613}

\bibitem[{{Lebouteiller} {et~al.}(2006){Lebouteiller}, {Kunth}, {Lequeux},
  {Aloisi}, {D{\'e}sert}, {H{\'e}brard}, {Lecavelier Des {\'E}tangs}, \&
  {Vidal-Madjar}}]{Leb2006}
{Lebouteiller}, V., {Kunth}, D., {Lequeux}, J., {et~al.} 2006, \aap, 459, 161,
  \dodoi{10.1051/0004-6361:20053161}

\bibitem[{{Lee} \& {Lee}(2014)}]{LeeLee2014}
{Lee}, J.~H., \& {Lee}, M.~G. 2014, \apj, 793, 134,
  \dodoi{10.1088/0004-637X/793/2/134}

\bibitem[{{Leitherer} {et~al.}(1992){Leitherer}, {Robert}, \&
  {Drissen}}]{Leitherer1992}
{Leitherer}, C., {Robert}, C., \& {Drissen}, L. 1992, \apj, 401, 596,
  \dodoi{10.1086/172089}

\bibitem[{{Li} {et~al.}(2004){Li}, {Kastner}, {Prigozhin}, {Schulz},
  {Feigelson}, \& {Getman}}]{Li2004}
{Li}, J., {Kastner}, J.~H., {Prigozhin}, G.~Y., {et~al.} 2004, \apj, 610, 1204,
  \dodoi{10.1086/421866}

\bibitem[{{Liu} {et~al.}(2013){Liu}, {Bregman}, {Bai}, {Justham}, \&
  {Crowther}}]{Liu2013}
{Liu}, J.-F., {Bregman}, J.~N., {Bai}, Y., {Justham}, S., \& {Crowther}, P.
  2013, \nat, 503, 500, \dodoi{10.1038/nature12762}

\bibitem[{{Lomax} {et~al.}(2015){Lomax}, {Naz{\'e}}, {Hoffman}, {Russell}, {De
  Becker}, {Corcoran}, {Davidson}, {Neilson}, {Owocki}, {Pittard}, \&
  {Pollock}}]{Lomax2015}
{Lomax}, J.~R., {Naz{\'e}}, Y., {Hoffman}, J.~L., {et~al.} 2015, \aap, 573,
  A43, \dodoi{10.1051/0004-6361/201424468}

\bibitem[{{Long} \& {White}(1980)}]{Long1980}
{Long}, K.~S., \& {White}, R.~L. 1980, \apj, 239, L65, \dodoi{10.1086/183293}

\bibitem[{{Long} {et~al.}(2010){Long}, {Blair}, {Winkler}, {Becker}, {Gaetz},
  {Ghavamian}, {Helfand}, {Hughes}, {Kirshner}, {Kuntz}, {McNeil}, {Pannuti},
  {Plucinsky}, {Saul}, {T{\"u}llmann}, \& {Williams}}]{Long2010}
{Long}, K.~S., {Blair}, W.~P., {Winkler}, P.~F., {et~al.} 2010, The
  Astrophysical Journal Supplement Series, 187, 495,
  \dodoi{10.1088/0067-0049/187/2/495}

\bibitem[{{Lucy}(1982)}]{Lucy1982}
{Lucy}, L.~B. 1982, \apj, 255, 286, \dodoi{10.1086/159827}

\bibitem[{{Lucy} \& {Solomon}(1970)}]{Lucy1970}
{Lucy}, L.~B., \& {Solomon}, P.~M. 1970, \apj, 159, 879, \dodoi{10.1086/150365}

\bibitem[{{Lucy} \& {White}(1980)}]{Lucy1980}
{Lucy}, L.~B., \& {White}, R.~L. 1980, \apj, 241, 300, \dodoi{10.1086/158342}

\bibitem[{{Luo} {et~al.}(1990){Luo}, {McCray}, \& {Mac Low}}]{Luo1990}
{Luo}, D., {McCray}, R., \& {Mac Low}, M.-M. 1990, \apj, 362, 267,
  \dodoi{10.1086/169263}

\bibitem[{{Maccarone} {et~al.}(2014){Maccarone}, {Lehmer}, {Leyder},
  {Antoniou}, {Hornschemeier}, {Ptak}, {Wik}, \& {Zezas}}]{Maccarone2014}
{Maccarone}, T.~J., {Lehmer}, B.~D., {Leyder}, J.~C., {et~al.} 2014, \mnras,
  439, 3064, \dodoi{10.1093/mnras/stu167}

\bibitem[{{Magrini} {et~al.}(2007){Magrini}, {V{\'\i}lchez}, {Mampaso},
  {Corradi}, \& {Leisy}}]{Magrini2007}
{Magrini}, L., {V{\'\i}lchez}, J.~M., {Mampaso}, A., {Corradi}, R.~L.~M., \&
  {Leisy}, P. 2007, \aap, 470, 865, \dodoi{10.1051/0004-6361:20077445}

\bibitem[{{Ma{\'{\i}}z-Apell{\'a}niz}
  {et~al.}(2004){Ma{\'{\i}}z-Apell{\'a}niz}, {P{\'e}rez}, \&
  {Mas-Hesse}}]{MA2004}
{Ma{\'{\i}}z-Apell{\'a}niz}, J., {P{\'e}rez}, E., \& {Mas-Hesse}, J.~M. 2004,
  \aj, 128, 1196, \dodoi{10.1086/422925}

\bibitem[{{Marchenko} {et~al.}(2003){Marchenko}, {Moffat}, {Ballereau},
  {Chauville}, {Zorec}, {Hill}, {Annuk}, {Corral}, {Demers}, {Eenens}, {Panov},
  {Seggewiss}, {Thomson}, \& {Villar-Sbaffi}}]{Marchenko2003}
{Marchenko}, S.~V., {Moffat}, A.~F.~J., {Ballereau}, D., {et~al.} 2003, \apj,
  596, 1295, \dodoi{10.1086/378154}

\bibitem[{{Marigo} {et~al.}(2017){Marigo}, {Girardi}, {Bressan}, {Rosenfield},
  {Aringer}, {Chen}, {Dussin}, {Nanni}, {Pastorelli}, {Rodrigues}, {Trabucchi},
  {Bladh}, {Dalcanton}, {Groenewegen}, {Montalb{\'a}n}, \& {Wood}}]{Marigo2017}
{Marigo}, P., {Girardi}, L., {Bressan}, A., {et~al.} 2017, \apj, 835, 77,
  \dodoi{10.3847/1538-4357/835/1/77}

\bibitem[{{Massey}(1998)}]{Massey1998Mv}
{Massey}, P. 1998, in Astronomical Society of the Pacific Conference Series,
  Vol. 142, The Stellar Initial Mass Function (38th Herstmonceux Conference),
  ed. G.~{Gilmore} \& D.~{Howell}, 17

\bibitem[{{Massey} {et~al.}(1995){Massey}, {Armandroff}, {Pyke}, {Patel}, \&
  {Wilson}}]{Massey1995}
{Massey}, P., {Armandroff}, T.~E., {Pyke}, R., {Patel}, K., \& {Wilson}, C.~D.
  1995, \aj, 110, 2715, \dodoi{10.1086/117725}

\bibitem[{{Massey} {et~al.}(1996){Massey}, {Bianchi}, {Hutchings}, \&
  {Stecher}}]{Massey1996}
{Massey}, P., {Bianchi}, L., {Hutchings}, J.~B., \& {Stecher}, T.~P. 1996,
  \apj, 469, 629, \dodoi{10.1086/177811}

\bibitem[{{Massey} \& {Conti}(1983)}]{Massey1983}
{Massey}, P., \& {Conti}, P.~S. 1983, \apj, 273, 576, \dodoi{10.1086/161393}

\bibitem[{{Massey} \& {Johnson}(1998)}]{Massey1998}
{Massey}, P., \& {Johnson}, O. 1998, \apj, 505, 793, \dodoi{10.1086/306199}

\bibitem[{{Massey} {et~al.}(2014){Massey}, {Neugent}, {Morrell}, \&
  {Hillier}}]{Massey2014}
{Massey}, P., {Neugent}, K.~F., {Morrell}, N., \& {Hillier}, D.~J. 2014, \apj,
  788, 83, \dodoi{10.1088/0004-637X/788/1/83}

\bibitem[{{Massey} {et~al.}(2016){Massey}, {Neugent}, \& {Smart}}]{Massey2016}
{Massey}, P., {Neugent}, K.~F., \& {Smart}, B.~M. 2016, \aj, 152, 62,
  \dodoi{10.3847/0004-6256/152/3/62}

\bibitem[{{Massey} {et~al.}(2006){Massey}, {Olsen}, {Hodge}, {Strong},
  {Jacoby}, {Schlingman}, \& {Smith}}]{Massey2006}
{Massey}, P., {Olsen}, K.~A.~G., {Hodge}, P.~W., {et~al.} 2006, \aj, 131, 2478,
  \dodoi{10.1086/503256}

\bibitem[{{Naz{\'e}} {et~al.}(2013){Naz{\'e}}, {Oskinova}, \&
  {Gosset}}]{Naze2013}
{Naz{\'e}}, Y., {Oskinova}, L.~M., \& {Gosset}, E. 2013, \apj, 763, 143,
  \dodoi{10.1088/0004-637X/763/2/143}

\bibitem[{{Naz{\'e}} {et~al.}(2011){Naz{\'e}}, {Broos}, {Oskinova}, {Townsley},
  {Cohen}, {Corcoran}, {Evans}, {Gagn{\'e}}, {Moffat}, {Pittard}, {Rauw},
  {ud-Doula}, \& {Walborn}}]{Naze2011}
{Naz{\'e}}, Y., {Broos}, P.~S., {Oskinova}, L., {et~al.} 2011, The
  Astrophysical Journal Supplement Series, 194, 7,
  \dodoi{10.1088/0067-0049/194/1/7}

\bibitem[{{Neugent} \& {Massey}(2011)}]{Neugent2011}
{Neugent}, K.~F., \& {Massey}, P. 2011, \apj, 733, 123,
  \dodoi{10.1088/0004-637X/733/2/123}

\bibitem[{{Neugent} \& {Massey}(2014)}]{Neugent2014}
---. 2014, \apj, 789, 10, \dodoi{10.1088/0004-637X/789/1/10}

\bibitem[{{Neugent} {et~al.}(2018){Neugent}, {Massey}, \&
  {Morrell}}]{Neugent2018}
{Neugent}, K.~F., {Massey}, P., \& {Morrell}, N. 2018, \apj, 863, 181,
  \dodoi{10.3847/1538-4357/aad17d}

\bibitem[{{Nugis} \& {Lamers}(2000)}]{Nugis2000}
{Nugis}, T., \& {Lamers}, H.~J.~G.~L.~M. 2000, \aap, 360, 227

\bibitem[{{Oskinova} {et~al.}(2004){Oskinova}, {Feldmeier}, \&
  {Hamann}}]{Oskinova2004}
{Oskinova}, L.~M., {Feldmeier}, A., \& {Hamann}, W.-R. 2004, \aap, 422, 675,
  \dodoi{10.1051/0004-6361:20047187}

\bibitem[{{Oskinova} {et~al.}(2007){Oskinova}, {Hamann}, \&
  {Feldmeier}}]{Oskinova2007}
{Oskinova}, L.~M., {Hamann}, W.-R., \& {Feldmeier}, A. 2007, \aap, 476, 1331,
  \dodoi{10.1051/0004-6361:20066377}

\bibitem[{{Oskinova} {et~al.}(2003){Oskinova}, {Ignace}, {Hamann}, {Pollock},
  \& {Brown}}]{Oskinova2003}
{Oskinova}, L.~M., {Ignace}, R., {Hamann}, W.~R., {Pollock}, A.~M.~T., \&
  {Brown}, J.~C. 2003, \aap, 402, 755, \dodoi{10.1051/0004-6361:20030300}

\bibitem[{{Owocki} {et~al.}(1988){Owocki}, {Castor}, \& {Rybicki}}]{Owocki1988}
{Owocki}, S.~P., {Castor}, J.~I., \& {Rybicki}, G.~B. 1988, \apj, 335, 914,
  \dodoi{10.1086/166977}

\bibitem[{{Park} {et~al.}(2006){Park}, {Kashyap}, {Siemiginowska}, {van Dyk},
  {Zezas}, {Heinke}, \& {Wargelin}}]{BEHR}
{Park}, T., {Kashyap}, V.~L., {Siemiginowska}, A., {et~al.} 2006, \apj, 652,
  610, \dodoi{10.1086/507406}

\bibitem[{{Pittard}(2007)}]{Pittard2007}
{Pittard}, J.~M. 2007, \apjl, 660, L141, \dodoi{10.1086/518365}

\bibitem[{{Pittard} \& {Corcoran}(2002)}]{Pittard2002}
{Pittard}, J.~M., \& {Corcoran}, M.~F. 2002, \aap, 383, 636,
  \dodoi{10.1051/0004-6361:20020025}

\bibitem[{{Pittard} \& {Dougherty}(2006)}]{Pittard2006}
{Pittard}, J.~M., \& {Dougherty}, S.~M. 2006, \mnras, 372, 801,
  \dodoi{10.1111/j.1365-2966.2006.10888.x}

\bibitem[{{Pittard} \& {Stevens}(1997)}]{Pittard1997}
{Pittard}, J.~M., \& {Stevens}, I.~R. 1997, \mnras, 292, 298,
  \dodoi{10.1093/mnras/292.2.298}

\bibitem[{{Pittard} \& {Stevens}(2002)}]{Pittard2002wind}
---. 2002, \aap, 388, L20, \dodoi{10.1051/0004-6361:20020583}

\bibitem[{{Pollock}(1987)}]{Pollock1987}
{Pollock}, A.~M.~T. 1987, \apj, 320, 283, \dodoi{10.1086/165539}

\bibitem[{{Pollock} \& {Corcoran}(2006)}]{Pollock2006}
{Pollock}, A.~M.~T., \& {Corcoran}, M.~F. 2006, \aap, 445, 1093,
  \dodoi{10.1051/0004-6361:20053496}

\bibitem[{{Pollock} {et~al.}(2005){Pollock}, {Corcoran}, {Stevens}, \&
  {Williams}}]{Pollock2005}
{Pollock}, A.~M.~T., {Corcoran}, M.~F., {Stevens}, I.~R., \& {Williams}, P.~M.
  2005, \apj, 629, 482, \dodoi{10.1086/431193}

\bibitem[{{Pollock} {et~al.}(2018){Pollock}, {Crowther}, {Tehrani}, {Broos}, \&
  {Townsley}}]{Pollock2018}
{Pollock}, A.~M.~T., {Crowther}, P.~A., {Tehrani}, K., {Broos}, P.~S., \&
  {Townsley}, L.~K. 2018, \mnras, 474, 3228, \dodoi{10.1093/mnras/stx2879}

\bibitem[{{Prestwich} {et~al.}(2007){Prestwich}, {Kilgard}, {Crowther},
  {Carpano}, {Pollock}, {Zezas}, {Saar}, {Roberts}, \& {Ward}}]{Prestwich2007}
{Prestwich}, A.~H., {Kilgard}, R., {Crowther}, P.~A., {et~al.} 2007, \apjl,
  669, L21, \dodoi{10.1086/523755}

\bibitem[{{Prilutskii} \& {Usov}(1976)}]{Pri1976}
{Prilutskii}, O.~F., \& {Usov}, V.~V. 1976, \azh, 53, 6

\bibitem[{{Puebla} {et~al.}(2016){Puebla}, {Hillier}, {Zsarg{\'o}}, {Cohen}, \&
  {Leutenegger}}]{Puebla2016}
{Puebla}, R.~E., {Hillier}, D.~J., {Zsarg{\'o}}, J., {Cohen}, D.~H., \&
  {Leutenegger}, M.~A. 2016, \mnras, 456, 2907, \dodoi{10.1093/mnras/stv2783}

\bibitem[{{Rauw} \& {Naz{\'e}}(2016)}]{Rauw2016}
{Rauw}, G., \& {Naz{\'e}}, Y. 2016, Advances in Space Research, 58, 761,
  \dodoi{10.1016/j.asr.2015.09.026}

\bibitem[{{Rauw} {et~al.}(2002){Rauw}, {Vreux}, {Stevens}, {Gosset}, {Sana},
  {Jamar}, \& {Mason}}]{Rauw2002}
{Rauw}, G., {Vreux}, J.~M., {Stevens}, I.~R., {et~al.} 2002, \aap, 388, 552,
  \dodoi{10.1051/0004-6361:20020523}

\bibitem[{{Rauw} {et~al.}(2004){Rauw}, {De Becker}, {Naz{\'e}}, {Crowther},
  {Gosset}, {Sana}, {van der Hucht}, {Vreux}, \& {Williams}}]{Rauw2004}
{Rauw}, G., {De Becker}, M., {Naz{\'e}}, Y., {et~al.} 2004, \aap, 420, L9,
  \dodoi{10.1051/0004-6361:20040150}

\bibitem[{{Rauw} {et~al.}(2015){Rauw}, {Naz{\'e}}, {Wright}, {Drake},
  {Guarcello}, {Prinja}, {Peck}, {Albacete Colombo}, {Herrero}, {Kobulnicky},
  {Sciortino}, \& {Vink}}]{Rauw2015}
{Rauw}, G., {Naz{\'e}}, Y., {Wright}, N.~J., {et~al.} 2015, The Astrophysical
  Journal Supplement Series, 221, 1, \dodoi{10.1088/0067-0049/221/1/1}

\bibitem[{{Read} {et~al.}(2011){Read}, {Rosen}, {Saxton}, \&
  {Ramirez}}]{Read2011}
{Read}, A.~M., {Rosen}, S.~R., {Saxton}, R.~D., \& {Ramirez}, J. 2011, \aap,
  534, A34, \dodoi{10.1051/0004-6361/201117525}

\bibitem[{{Sana} {et~al.}(2006){Sana}, {Rauw}, {Naz{\'e}}, {Gosset}, \&
  {Vreux}}]{Sana2006}
{Sana}, H., {Rauw}, G., {Naz{\'e}}, Y., {Gosset}, E., \& {Vreux}, J.~M. 2006,
  \mnras, 372, 661, \dodoi{10.1111/j.1365-2966.2006.10847.x}

\bibitem[{{Sana} {et~al.}(2012){Sana}, {de Mink}, {de Koter}, {Langer},
  {Evans}, {Gieles}, {Gosset}, {Izzard}, {Le Bouquin}, \&
  {Schneider}}]{Sana2012}
{Sana}, H., {de Mink}, S.~E., {de Koter}, A., {et~al.} 2012, Science, 337, 444,
  \dodoi{10.1126/science.1223344}

\bibitem[{{Sander} {et~al.}(2012){Sander}, {Hamann}, \& {Todt}}]{Sander2012}
{Sander}, A., {Hamann}, W.~R., \& {Todt}, H. 2012, \aap, 540, A144,
  \dodoi{10.1051/0004-6361/201117830}

\bibitem[{{Sciortino} {et~al.}(1990){Sciortino}, {Vaiana}, {Harnden},
  {Ramella}, {Morossi}, {Rosner}, \& {Schmitt}}]{Scior1990}
{Sciortino}, S., {Vaiana}, G.~S., {Harnden}, F.~R., J., {et~al.} 1990, \apj,
  361, 621, \dodoi{10.1086/169225}

\bibitem[{{Seward} \& {Chlebowski}(1982)}]{Seward1982}
{Seward}, F.~D., \& {Chlebowski}, T. 1982, \apj, 256, 530,
  \dodoi{10.1086/159929}

\bibitem[{{Silverman} \& {Filippenko}(2008)}]{Silverman2008}
{Silverman}, J.~M., \& {Filippenko}, A.~V. 2008, \apjl, 678, L17,
  \dodoi{10.1086/588096}

\bibitem[{{Smith}(2014)}]{Smith2014}
{Smith}, N. 2014, \araa, 52, 487, \dodoi{10.1146/annurev-astro-081913-040025}

\bibitem[{{Smith} {et~al.}(2001){Smith}, {Brickhouse}, {Liedahl}, \&
  {Raymond}}]{Smith2001}
{Smith}, R.~K., {Brickhouse}, N.~S., {Liedahl}, D.~A., \& {Raymond}, J.~C.
  2001, \apj, 556, L91, \dodoi{10.1086/322992}

\bibitem[{{Stevens} {et~al.}(1992){Stevens}, {Blondin}, \&
  {Pollock}}]{Stevens1992}
{Stevens}, I.~R., {Blondin}, J.~M., \& {Pollock}, A.~M.~T. 1992, \apj, 386,
  265, \dodoi{10.1086/171013}

\bibitem[{{Stevens} {et~al.}(1996){Stevens}, {Corcoran}, {Willis}, {Skinner},
  {Pollock}, {Nagase}, \& {Koyama}}]{Stevens1996}
{Stevens}, I.~R., {Corcoran}, M.~F., {Willis}, A.~J., {et~al.} 1996, \mnras,
  283, 589, \dodoi{10.1093/mnras/283.2.589}

\bibitem[{{Str{\"u}der} {et~al.}(2001){Str{\"u}der}, {Briel}, {Dennerl},
  {Hartmann}, {Kendziorra}, {Meidinger}, {Pfeffermann}, {Reppin}, {Aschenbach},
  {Bornemann}, {Br{\"a}uninger}, {Burkert}, {Elender}, {Freyberg}, {Haberl},
  {Hartner}, {Heuschmann}, {Hippmann}, {Kastelic}, {Kemmer}, {Kettenring},
  {Kink}, {Krause}, {M{\"u}ller}, {Oppitz}, {Pietsch}, {Popp}, {Predehl},
  {Read}, {Stephan}, {St{\"o}tter}, {Tr{\"u}mper}, {Holl}, {Kemmer}, {Soltau},
  {St{\"o}tter}, {Weber}, {Weichert}, {von Zanthier}, {Carathanassis}, {Lutz},
  {Richter}, {Solc}, {B{\"o}ttcher}, {Kuster}, {Staubert}, {Abbey}, {Holland},
  {Turner}, {Balasini}, {Bignami}, {La Palombara}, {Villa}, {Buttler},
  {Gianini}, {Lain{\'e}}, {Lumb}, \& {Dhez}}]{Struder2001}
{Str{\"u}der}, L., {Briel}, U., {Dennerl}, K., {et~al.} 2001, \aap, 365, L18,
  \dodoi{10.1051/0004-6361:20000066}

\bibitem[{{Sugawara} {et~al.}(2015){Sugawara}, {Maeda}, {Tsuboi}, {Hamaguchi},
  {Corcoran}, {Pollock}, {Moffat}, {Williams}, {Dougherty}, \&
  {Pittard}}]{Sugawara2015}
{Sugawara}, Y., {Maeda}, Y., {Tsuboi}, Y., {et~al.} 2015, Publications of the
  Astronomical Society of Japan, 67, 121, \dodoi{10.1093/pasj/psv099}

\bibitem[{{Sundqvist} {et~al.}(2011){Sundqvist}, {Puls}, {Feldmeier}, \&
  {Owocki}}]{Sundqvist2011}
{Sundqvist}, J.~O., {Puls}, J., {Feldmeier}, A., \& {Owocki}, S.~P. 2011, \aap,
  528, A64, \dodoi{10.1051/0004-6361/201015771}

\bibitem[{{Tehrani} {et~al.}(2019){Tehrani}, {Crowther}, {Bestenlehner},
  {Littlefair}, {Pollock}, {Parker}, \& {Schnurr}}]{Tehrani2019}
{Tehrani}, K.~A., {Crowther}, P.~A., {Bestenlehner}, J.~M., {et~al.} 2019,
  \mnras, 484, 2692, \dodoi{10.1093/mnras/stz147}

\bibitem[{{Torres} {et~al.}(1986){Torres}, {Conti}, \& {Massey}}]{Torres1986}
{Torres}, A.~V., {Conti}, P.~S., \& {Massey}, P. 1986, \apj, 300, 379,
  \dodoi{10.1086/163811}

\bibitem[{{T{\"u}llmann} {et~al.}(2008){T{\"u}llmann}, {Gaetz}, {Plucinsky},
  {Long}, {Hughes}, {Blair}, {Winkler}, {Pannuti}, {Breitschwerdt}, \&
  {Ghavamian}}]{Tullmann2008}
{T{\"u}llmann}, R., {Gaetz}, T.~J., {Plucinsky}, P.~P., {et~al.} 2008, \apj,
  685, 919, \dodoi{10.1086/591019}

\bibitem[{{T{\"u}llmann} {et~al.}(2011){T{\"u}llmann}, {Gaetz}, {Plucinsky},
  {Kuntz}, {Williams}, {Pietsch}, {Haberl}, {Long}, {Blair}, {Sasaki},
  {Winkler}, {Challis}, {Pannuti}, {Edgar}, {Helfand}, {Hughes}, {Kirshner},
  {Mazeh}, \& {Shporer}}]{Tullmann2011}
---. 2011, \apjs, 193, 31, \dodoi{10.1088/0067-0049/193/2/31}

\bibitem[{{Turner} {et~al.}(2001){Turner}, {Abbey}, {Arnaud}, {Balasini},
  {Barbera}, {Belsole}, {Bennie}, {Bernard}, {Bignami}, {Boer}, {Briel},
  {Butler}, {Cara}, {Chabaud}, {Cole}, {Collura}, {Conte}, {Cros}, {Denby},
  {Dhez}, {Di Coco}, {Dowson}, {Ferrando}, {Ghizzardi}, {Gianotti}, {Goodall},
  {Gretton}, {Griffiths}, {Hainaut}, {Hochedez}, {Holland}, {Jourdain},
  {Kendziorra}, {Lagostina}, {Laine}, {La Palombara}, {Lortholary}, {Lumb},
  {Marty}, {Molendi}, {Pigot}, {Poindron}, {Pounds}, {Reeves}, {Reppin},
  {Rothenflug}, {Salvetat}, {Sauvageot}, {Schmitt}, {Sembay}, {Short},
  {Spragg}, {Stephen}, {Str{\"u}der}, {Tiengo}, {Trifoglio}, {Tr{\"u}mper},
  {Vercellone}, {Vigroux}, {Villa}, {Ward}, {Whitehead}, \&
  {Zonca}}]{Turner2001}
{Turner}, M.~J.~L., {Abbey}, A., {Arnaud}, M., {et~al.} 2001, \aap, 365, L27,
  \dodoi{10.1051/0004-6361:20000087}

\bibitem[{{Usov}(1992)}]{Usov1992}
{Usov}, V.~V. 1992, \apj, 389, 635, \dodoi{10.1086/171236}

\bibitem[{{van den Heuvel}(2019)}]{vdh2019}
{van den Heuvel}, E. P.~J. 2019, arXiv e-prints, arXiv:1901.06939.
\newblock \doarXiv{1901.06939}

\bibitem[{{van den Heuvel} {et~al.}(2017){van den Heuvel}, {Portegies Zwart},
  \& {de Mink}}]{vdh2017}
{van den Heuvel}, E.~P.~J., {Portegies Zwart}, S.~F., \& {de Mink}, S.~E. 2017,
  \mnras, 471, 4256, \dodoi{10.1093/mnras/stx1430}

\bibitem[{{van Kerkwijk} {et~al.}(1996){van Kerkwijk}, {Geballe}, {King}, {van
  der Klis}, \& {van Paradijs}}]{vanK1996}
{van Kerkwijk}, M.~H., {Geballe}, T.~R., {King}, D.~L., {van der Klis}, M., \&
  {van Paradijs}, J. 1996, \aap, 314, 521

\bibitem[{{Vilchez} {et~al.}(1988){Vilchez}, {Pagel}, {Diaz}, {Terlevich}, \&
  {Edmunds}}]{Vilchez1988}
{Vilchez}, J.~M., {Pagel}, B.~E.~J., {Diaz}, A.~I., {Terlevich}, E., \&
  {Edmunds}, M.~G. 1988, \mnras, 235, 633, \dodoi{10.1093/mnras/235.3.633}

\bibitem[{{Vink} {et~al.}(2001){Vink}, {de Koter}, \& {Lamers}}]{Vink2001}
{Vink}, J.~S., {de Koter}, A., \& {Lamers}, H.~J.~G.~L.~M. 2001, \aap, 369,
  574, \dodoi{10.1051/0004-6361:20010127}

\bibitem[{{West} {et~al.}(2018){West}, {Lehmer}, {Wik}, {Yang}, {Walton},
  {Antoniou}, {Haberl}, {Hornschemeier}, {Maccarone}, {Plucinsky}, {Ptak},
  {Williams}, {Vulic}, {Yukita}, \& {Zezas}}]{West2018}
{West}, L.~A., {Lehmer}, B.~D., {Wik}, D., {et~al.} 2018, \apj, 869, 111,
  \dodoi{10.3847/1538-4357/aaec6b}

\bibitem[{{White} \& {Long}(1986)}]{White1986}
{White}, R.~L., \& {Long}, K.~S. 1986, \apj, 310, 832, \dodoi{10.1086/164736}

\bibitem[{{Williams} {et~al.}(2014){Williams}, {Lang}, {Dalcanton}, {Dolphin},
  {Weisz}, {Bell}, {Bianchi}, {Byler}, {Gilbert}, {Girardi}, {Gordon},
  {Gregersen}, {Johnson}, {Kalirai}, {Lauer}, {Monachesi}, {Rosenfield},
  {Seth}, \& {Skillman}}]{Williams2014}
{Williams}, B.~F., {Lang}, D., {Dalcanton}, J.~J., {et~al.} 2014, \apjs, 215,
  9, \dodoi{10.1088/0067-0049/215/1/9}

\bibitem[{{Williams} {et~al.}(2015){Williams}, {Wold}, {Haberl}, {Garofali},
  {Blair}, {Gaetz}, {Kuntz}, {Long}, {Pannuti}, {Pietsch}, {Plucinsky}, \&
  {Winkler}}]{Williams2015}
{Williams}, B.~F., {Wold}, B., {Haberl}, F., {et~al.} 2015, \apjs, 218, 9,
  \dodoi{10.1088/0067-0049/218/1/9}

\bibitem[{{Williams} {et~al.}(2012){Williams}, {van der Hucht}, {van Wyk},
  {Marang}, {Whitelock}, {Bouchet}, \& {Setia Gunawan}}]{Williams2012}
{Williams}, P.~M., {van der Hucht}, K.~A., {van Wyk}, F., {et~al.} 2012,
  \mnras, 420, 2526, \dodoi{10.1111/j.1365-2966.2011.20218.x}

\bibitem[{{Yang} {et~al.}(1996){Yang}, {Chu}, {Skillman}, \&
  {Terlevich}}]{Yang1996}
{Yang}, H., {Chu}, Y.-H., {Skillman}, E.~D., \& {Terlevich}, R. 1996, \aj, 112,
  146, \dodoi{10.1086/117995}

\bibitem[{{Zhekov} {et~al.}(2011){Zhekov}, {Gagn{\'e}}, \&
  {Skinner}}]{Zhekov2011}
{Zhekov}, S.~A., {Gagn{\'e}}, M., \& {Skinner}, S.~L. 2011, \apj, 727, L17,
  \dodoi{10.1088/2041-8205/727/1/L17}

\bibitem[{{Zhekov} {et~al.}(2014){Zhekov}, {Tomov}, {Gawronski}, {Georgiev},
  {Borissova}, {Kurtev}, {Gagn{\'e}}, \& {Hajduk}}]{Zhekov2014WN8}
{Zhekov}, S.~A., {Tomov}, T., {Gawronski}, M.~P., {et~al.} 2014, \mnras, 445,
  1663, \dodoi{10.1093/mnras/stu1880}

\end{thebibliography}

%%%%%%%%%%%%%%%%%%%%%%%%%%%%%%%%%%%%%%%%%%%%%%%%%%

% Don't change these lines
%\bsp	% typesetting comment
%\label{lastpage}
\end{document}